\documentclass[letterpaper,twocolumn,10pt]{article}

\usepackage[usenames,dvipsnames]{xcolor}

\usepackage{sty/usenix2019_v3}
\usepackage[square,comma,numbers,sort&compress]{natbib}
\usepackage{xtab}

\usepackage{amssymb}
\usepackage{pifont}

\usepackage{amsmath,amsopn}
\usepackage{subfigure}
\usepackage{endnotes}
\usepackage{microtype}
\usepackage{xspace}
\usepackage{graphicx}
\usepackage{fancyvrb}
\usepackage{multirow}
\usepackage{array}
\usepackage{underscore}
\usepackage{relsize}
\usepackage{pbox}
\usepackage{xstring}
\usepackage{tikz}
\usepackage[linesnumbered,lined,ruled]{algorithm2e}
\usepackage{fp}
\usepackage{siunitx}
\usepackage{flushend}
\usepackage{balance}
\usepackage{multicol}
\usepackage{hyperref}
\usepackage{booktabs}
\usepackage{makecell}
\usepackage{enumitem}

\usepackage{color, colortbl}

\makeatletter
\def\mdseries@tt{m}
\makeatother
\usepackage{minted}

\fvset{fontsize=\scriptsize,xleftmargin=8pt,numbers=left,numbersep=5pt}

\makeatletter
\def\PY@reset{\let\PY@it=\relax \let\PY@bf=\relax%
    \let\PY@ul=\relax \let\PY@tc=\relax%
    \let\PY@bc=\relax \let\PY@ff=\relax}
\def\PY@tok#1{\csname PY@tok@#1\endcsname}
\def\PY@toks#1+{\ifx\relax#1\empty\else%
    \PY@tok{#1}\expandafter\PY@toks\fi}
\def\PY@do#1{\PY@bc{\PY@tc{\PY@ul{%
    \PY@it{\PY@bf{\PY@ff{#1}}}}}}}
\def\PY#1#2{\PY@reset\PY@toks#1+\relax+\PY@do{#2}}

\expandafter\def\csname PY@tok@gd\endcsname{\def\PY@tc##1{\textcolor[rgb]{0.63,0.00,0.00}{##1}}}
\expandafter\def\csname PY@tok@gu\endcsname{\let\PY@bf=\textbf\def\PY@tc##1{\textcolor[rgb]{0.50,0.00,0.50}{##1}}}
\expandafter\def\csname PY@tok@gt\endcsname{\def\PY@tc##1{\textcolor[rgb]{0.00,0.27,0.87}{##1}}}
\expandafter\def\csname PY@tok@gs\endcsname{\let\PY@bf=\textbf}
\expandafter\def\csname PY@tok@gr\endcsname{\def\PY@tc##1{\textcolor[rgb]{1.00,0.00,0.00}{##1}}}
\expandafter\def\csname PY@tok@cm\endcsname{\let\PY@it=\textit\def\PY@tc##1{\textcolor[rgb]{0.25,0.50,0.50}{##1}}}
\expandafter\def\csname PY@tok@vg\endcsname{\def\PY@tc##1{\textcolor[rgb]{0.10,0.09,0.49}{##1}}}
\expandafter\def\csname PY@tok@vi\endcsname{\def\PY@tc##1{\textcolor[rgb]{0.10,0.09,0.49}{##1}}}
\expandafter\def\csname PY@tok@vm\endcsname{\def\PY@tc##1{\textcolor[rgb]{0.10,0.09,0.49}{##1}}}
\expandafter\def\csname PY@tok@mh\endcsname{\def\PY@tc##1{\textcolor[rgb]{0.40,0.40,0.40}{##1}}}
\expandafter\def\csname PY@tok@cs\endcsname{\let\PY@it=\textit\def\PY@tc##1{\textcolor[rgb]{0.25,0.50,0.50}{##1}}}
\expandafter\def\csname PY@tok@ge\endcsname{\let\PY@it=\textit}
\expandafter\def\csname PY@tok@vc\endcsname{\def\PY@tc##1{\textcolor[rgb]{0.10,0.09,0.49}{##1}}}
\expandafter\def\csname PY@tok@il\endcsname{\def\PY@tc##1{\textcolor[rgb]{0.40,0.40,0.40}{##1}}}
\expandafter\def\csname PY@tok@go\endcsname{\def\PY@tc##1{\textcolor[rgb]{0.53,0.53,0.53}{##1}}}
\expandafter\def\csname PY@tok@cp\endcsname{\def\PY@tc##1{\textcolor[rgb]{0.74,0.48,0.00}{##1}}}
\expandafter\def\csname PY@tok@gi\endcsname{\def\PY@tc##1{\textcolor[rgb]{0.00,0.63,0.00}{##1}}}
\expandafter\def\csname PY@tok@gh\endcsname{\let\PY@bf=\textbf\def\PY@tc##1{\textcolor[rgb]{0.00,0.00,0.50}{##1}}}
\expandafter\def\csname PY@tok@ni\endcsname{\let\PY@bf=\textbf\def\PY@tc##1{\textcolor[rgb]{0.60,0.60,0.60}{##1}}}
\expandafter\def\csname PY@tok@nl\endcsname{\def\PY@tc##1{\textcolor[rgb]{0.63,0.63,0.00}{##1}}}
\expandafter\def\csname PY@tok@nn\endcsname{\let\PY@bf=\textbf\def\PY@tc##1{\textcolor[rgb]{0.00,0.00,1.00}{##1}}}
\expandafter\def\csname PY@tok@no\endcsname{\def\PY@tc##1{\textcolor[rgb]{0.53,0.00,0.00}{##1}}}
\expandafter\def\csname PY@tok@na\endcsname{\def\PY@tc##1{\textcolor[rgb]{0.49,0.56,0.16}{##1}}}
\expandafter\def\csname PY@tok@nb\endcsname{\def\PY@tc##1{\textcolor[rgb]{0.00,0.50,0.00}{##1}}}
\expandafter\def\csname PY@tok@nc\endcsname{\let\PY@bf=\textbf\def\PY@tc##1{\textcolor[rgb]{0.00,0.00,1.00}{##1}}}
\expandafter\def\csname PY@tok@nd\endcsname{\def\PY@tc##1{\textcolor[rgb]{0.67,0.13,1.00}{##1}}}
\expandafter\def\csname PY@tok@ne\endcsname{\let\PY@bf=\textbf\def\PY@tc##1{\textcolor[rgb]{0.82,0.25,0.23}{##1}}}
\expandafter\def\csname PY@tok@nf\endcsname{\def\PY@tc##1{\textcolor[rgb]{0.00,0.00,1.00}{##1}}}
\expandafter\def\csname PY@tok@si\endcsname{\let\PY@bf=\textbf\def\PY@tc##1{\textcolor[rgb]{0.73,0.40,0.53}{##1}}}
\expandafter\def\csname PY@tok@s2\endcsname{\def\PY@tc##1{\textcolor[rgb]{0.73,0.13,0.13}{##1}}}
\expandafter\def\csname PY@tok@nt\endcsname{\let\PY@bf=\textbf\def\PY@tc##1{\textcolor[rgb]{0.00,0.50,0.00}{##1}}}
\expandafter\def\csname PY@tok@nv\endcsname{\def\PY@tc##1{\textcolor[rgb]{0.10,0.09,0.49}{##1}}}
\expandafter\def\csname PY@tok@s1\endcsname{\def\PY@tc##1{\textcolor[rgb]{0.73,0.13,0.13}{##1}}}
\expandafter\def\csname PY@tok@dl\endcsname{\def\PY@tc##1{\textcolor[rgb]{0.73,0.13,0.13}{##1}}}
\expandafter\def\csname PY@tok@ch\endcsname{\let\PY@it=\textit\def\PY@tc##1{\textcolor[rgb]{0.25,0.50,0.50}{##1}}}
\expandafter\def\csname PY@tok@m\endcsname{\def\PY@tc##1{\textcolor[rgb]{0.40,0.40,0.40}{##1}}}
\expandafter\def\csname PY@tok@gp\endcsname{\let\PY@bf=\textbf\def\PY@tc##1{\textcolor[rgb]{0.00,0.00,0.50}{##1}}}
\expandafter\def\csname PY@tok@sh\endcsname{\def\PY@tc##1{\textcolor[rgb]{0.73,0.13,0.13}{##1}}}
\expandafter\def\csname PY@tok@ow\endcsname{\let\PY@bf=\textbf\def\PY@tc##1{\textcolor[rgb]{0.67,0.13,1.00}{##1}}}
\expandafter\def\csname PY@tok@sx\endcsname{\def\PY@tc##1{\textcolor[rgb]{0.00,0.50,0.00}{##1}}}
\expandafter\def\csname PY@tok@bp\endcsname{\def\PY@tc##1{\textcolor[rgb]{0.00,0.50,0.00}{##1}}}
\expandafter\def\csname PY@tok@c1\endcsname{\let\PY@it=\textit\def\PY@tc##1{\textcolor[rgb]{0.25,0.50,0.50}{##1}}}
\expandafter\def\csname PY@tok@fm\endcsname{\def\PY@tc##1{\textcolor[rgb]{0.00,0.00,1.00}{##1}}}
\expandafter\def\csname PY@tok@o\endcsname{\def\PY@tc##1{\textcolor[rgb]{0.40,0.40,0.40}{##1}}}
\expandafter\def\csname PY@tok@kc\endcsname{\let\PY@bf=\textbf\def\PY@tc##1{\textcolor[rgb]{0.00,0.50,0.00}{##1}}}
\expandafter\def\csname PY@tok@c\endcsname{\let\PY@it=\textit\def\PY@tc##1{\textcolor[rgb]{0.25,0.50,0.50}{##1}}}
\expandafter\def\csname PY@tok@mf\endcsname{\def\PY@tc##1{\textcolor[rgb]{0.40,0.40,0.40}{##1}}}
\expandafter\def\csname PY@tok@err\endcsname{\def\PY@bc##1{\setlength{\fboxsep}{0pt}\fcolorbox[rgb]{1.00,0.00,0.00}{1,1,1}{\strut ##1}}}
\expandafter\def\csname PY@tok@mb\endcsname{\def\PY@tc##1{\textcolor[rgb]{0.40,0.40,0.40}{##1}}}
\expandafter\def\csname PY@tok@ss\endcsname{\def\PY@tc##1{\textcolor[rgb]{0.10,0.09,0.49}{##1}}}
\expandafter\def\csname PY@tok@sr\endcsname{\def\PY@tc##1{\textcolor[rgb]{0.73,0.40,0.53}{##1}}}
\expandafter\def\csname PY@tok@mo\endcsname{\def\PY@tc##1{\textcolor[rgb]{0.40,0.40,0.40}{##1}}}
\expandafter\def\csname PY@tok@kd\endcsname{\let\PY@bf=\textbf\def\PY@tc##1{\textcolor[rgb]{0.00,0.50,0.00}{##1}}}
\expandafter\def\csname PY@tok@mi\endcsname{\def\PY@tc##1{\textcolor[rgb]{0.40,0.40,0.40}{##1}}}
\expandafter\def\csname PY@tok@kn\endcsname{\let\PY@bf=\textbf\def\PY@tc##1{\textcolor[rgb]{0.00,0.50,0.00}{##1}}}
\expandafter\def\csname PY@tok@cpf\endcsname{\let\PY@it=\textit\def\PY@tc##1{\textcolor[rgb]{0.25,0.50,0.50}{##1}}}
\expandafter\def\csname PY@tok@kr\endcsname{\let\PY@bf=\textbf\def\PY@tc##1{\textcolor[rgb]{0.00,0.50,0.00}{##1}}}
\expandafter\def\csname PY@tok@s\endcsname{\def\PY@tc##1{\textcolor[rgb]{0.73,0.13,0.13}{##1}}}
\expandafter\def\csname PY@tok@kp\endcsname{\def\PY@tc##1{\textcolor[rgb]{0.00,0.50,0.00}{##1}}}
\expandafter\def\csname PY@tok@w\endcsname{\def\PY@tc##1{\textcolor[rgb]{0.73,0.73,0.73}{##1}}}
\expandafter\def\csname PY@tok@kt\endcsname{\def\PY@tc##1{\textcolor[rgb]{0.69,0.00,0.25}{##1}}}
\expandafter\def\csname PY@tok@sc\endcsname{\def\PY@tc##1{\textcolor[rgb]{0.73,0.13,0.13}{##1}}}
\expandafter\def\csname PY@tok@sb\endcsname{\def\PY@tc##1{\textcolor[rgb]{0.73,0.13,0.13}{##1}}}
\expandafter\def\csname PY@tok@sa\endcsname{\def\PY@tc##1{\textcolor[rgb]{0.73,0.13,0.13}{##1}}}
\expandafter\def\csname PY@tok@k\endcsname{\let\PY@bf=\textbf\def\PY@tc##1{\textcolor[rgb]{0.00,0.50,0.00}{##1}}}
\expandafter\def\csname PY@tok@se\endcsname{\let\PY@bf=\textbf\def\PY@tc##1{\textcolor[rgb]{0.73,0.40,0.13}{##1}}}
\expandafter\def\csname PY@tok@sd\endcsname{\let\PY@it=\textit\def\PY@tc##1{\textcolor[rgb]{0.73,0.13,0.13}{##1}}}

\makeatother

\def\Snospace~{\S{}}

\newcommand{\cc}[1]{\mbox{\smaller[0.5]\texttt{#1}}}
\newcommand{\PP}[1]{
\vspace{2px}
\noindent{\bf \IfEndWith{#1}{.}{#1}{#1.}}
}

\usepackage{pifont}
\newcommand{\cmark}{\ding{51}}%
\definecolor{Gray}{gray}{0.9}

\newcommand{\boxbeg}{
\vspace{2px}
\noindent\begin{tabular}{|l|}\hline
\begin{minipage}{3.2in}
\vspace{2px}
\noindent
}

\newcommand{\boxend}{
\vspace{2px}
\end{minipage}\\ \hline
\end{tabular}
\vspace{-10pt}
}

\newcommand{\sys}{\mbox{\textsc{BugStone}}\xspace}
\newcommand{\hpb}{\mbox{\textsc{RPB}s}\xspace}
\newcommand{\graycc}{\cellcolor{gray!20}}

\newcommand{\QW}[1]{\textcolor{red}{[QW: #1]}}

\newcommand{\XXX}[1]{\textcolor{red}{XXX: #1}}
\newcommand{\ignore}[1]{}

\newcounter{notesec}[section]
\newcommand{\thenoteee}{\thesection.\arabic{notesec}}
\newcommand{\yue}[1]{\refstepcounter{notesec}{\bf\textcolor{blue}{$\ll$Yue~\thenoteee: {\sf #1}$\gg$}}}

\fvset{fontsize=\scriptsize,xleftmargin=8pt,numbers=left,numbersep=5pt}

\input{code/fmt}

\usepackage{array}    
\usepackage{ragged2e} 
\newcolumntype{Z}[1]{>{\justifying\arraybackslash}p{#1}}

\def\Snospace~{\S{}}

\SetCommentSty{mycommfont}

\newif\ifdraft\drafttrue
\newif\ifnotes\notestrue
\ifdraft\else\notesfalse\fi

\input{glyphtounicode}
\newcolumntype{R}[1]{>{\raggedleft\let\newline\\\arraybackslash\hspace{0pt}}p{#1}}

\newcommand{\squishlist}{
\begin{itemize}[noitemsep,nolistsep]
  \setlength{\itemsep}{-0pt}
}
\newcommand{\squishend}{
  \end{itemize}
}

\usepackage{tikz}

\usepackage{xstring}

\begin{document}

\title{One Bug, Hundreds Behind: LLMs for Large-Scale Bug Discovery}

\ifdefined\DRAFT
 \pagestyle{fancyplain}
 \lhead{Rev.~\therev}
 \rhead{\thedate}
 \cfoot{\thepage\ of \pageref{LastPage}}
\fi



\author{
 {\rm Qiushi Wu$^{1}$,\;
 Yue Xiao$^{2}$,\;
 Dhilung Kirat$^{1}$,\;
 Kevin Eykholt$^{1}$,\;
 Jiyong Jang$^{1}$,\;
 Douglas Lee Schales$^{1}$}\\
$^1$IBM Research, 
$^2$William \& Mary\\
}

\date{}

\maketitle

\sloppy
\begin{abstract}







Managing large programs and fixing bugs is a challenging task that demands substantial time and manual effort. In practice, when a bug is found, it is reported to the project maintainers, who work with the reporter to fix it. Once the bug is resolved, the issue is marked as closed. However, across the program, there are often similar code segments, which may also contain the bug, but were initially missed during discovery. Finding and fixing each recurring bug instance individually is labor intensive. Even more concerning, bug reports can inadvertently widen the attack surface as they provide attackers with an exploitable pattern that may be unresolved in other parts of the program.

In this paper, we explore these Recurring Pattern Bugs (RPBs) that appear repeatedly across various code segments of a program or even in different programs, stemming from a same root cause, but are unresolved, even when a fix is available. Our investigation reveals that RPBs are not only prevalent, but also have the potential to significantly compromise the security of software programs. Various static analyzers exist for finding specific bug patterns but require significant engineering effort and don't generalize well beyond their predefined template. To tackle RPBs and enhance project maintenance efficiency, this paper introduces \sys, a program analysis system empowered by LLVM and a Large Language Model (LLM). The key observation is that many RPBs have a single instance patched, which can be leveraged to identify a consistent error pattern, such as a specific API misuse. By examining the entire program for other instances of this pattern, it is possible to identify similar sections of code that are vulnerable to the same type of bug. Starting with 135 unique seed RPBs, \sys was able to identify more than 22K new potential issues in the Linux kernel. Manual analysis of 400 of these findings confirmed that 246 are valid issues. These detected issues include invalid pointer dereferences, resource leaks, type errors, performance issues, and others, threatening the security and stability of the program. We also create a dataset from over 1.9K security bugs reported by 23 recent top-tier conference works targeting bug identification in the Linux kernel. We manually annotate the dataset, identify 80 recurring patterns and 850 corresponding fixes, and evaluate \sys on this dataset across six different LLMs and six prompt configurations. Even with a cost-efficient model choice, \sys achieved 92.2\% precision and 79.1\% pairwise accuracy, demonstrating its capability to accurately identify RPBs in real-world programs.

\end{abstract}

\section{Introduction}
\label{s:intro}



Maintaining large programs requires considerable manual labor for debugging, enhancing performance, upgrading features, and so on. 
For example, a recent study~\cite{baitowards} reveals that, in recent years, the Linux kernel received over 80,000 commits yearly, with around 10,000 for bug fixes, underscoring substantial maintenance efforts.
Over the years, the escalation in program size and complexity has worsened this situation and increased the volume of bug reports and the need for maintenance. 
For instance, using Cloc~\cite{Cloc} to measure the size of the code base, the Linux kernel has grown from 5 million lines of code in version 2.6.12 to 28 million in version 6.8, reflecting the significant growth in the program scale.

Given the considerable size of the code bases, large projects often adopt the following two maintenance strategies.
First, modular maintenance, which assigns one or several maintainers to manage a specific submodule of the larger program, allowing for more focused and effective management.
Second, open reporting, which encourages all users to contribute by reporting bugs or submitting patches whenever they encounter issues, promoting a collaborative and inclusive maintenance process.

\PP{Inherent security challenges in modular and collaborative maintenance strategies} While this maintenance approach is highly effective for independent modules, these strategies suffer from the following shortcomings and raise potential security concerns.
First, updates to API functions may not be promptly or adequately integrated into all downstream or dependent 
modules. This is a well-known issue in supply chains~\cite{williams2024research}, where maintainers oversee various modules, and might not be fully aware of or understand the implication of API changes in a timely manner.
Second, similar bugs may arise across different modules due to recurring oversights by multiple contributors and the strategy of modular maintenance.
Users typically focus on addressing the specific issue they encounter, which allows similar problems to remain in other modules or code areas. As such, duplicate effort is wasted repeatedly identifying, reporting, reviewing, testing, and fixing similar recurring bugs.
For instance, recent studies~\cite{he2023one, lin2023detecting,HERO} indicate that a single API function can give rise to hundreds of similar bugs, each subsequently addressed by individual patches.
Finally, the most critical issue is that bug fixes for a single instance of a recurring issue can unintentionally help attackers find and exploit similar latent vulnerabilities throughout the code base.
Therefore, such problems will not only exacerbate maintenance challenges but also cause security issues~\cite{wu2022aware}.

To address such maintenance and security issues in large software projects efficiently, modern strategies concentrate on three core approaches.
First, bug report classification and prioritization~\cite{wu2020precisely,wu2022aware,wang2020machine,zhou2017automated} enable maintainers to swiftly address the critical issues, thereby reducing the lifespan of vulnerabilities within the software.
Second, identifying duplicate bug reports~\cite{mu2022depth,he2020duplicate,kukkar2020duplicate} enhances maintenance efficiency by preventing redundant work.
And third, bug detection harnesses both static and dynamic analysis techniques, including rule-based methods~\cite{nielebock2020cooperative}, statistical approaches~\cite{lu2019detecting}, and fuzzing~\cite{kim2020hfl}, to proactively identify and resolve bugs from diverse perspectives, thereby strengthening the reliability and security of software.

However, with independent bug detection approaches, the root cause, \textit{unresolved recurring bugs}, still cannot be properly addressed.
Given the size of the current code bases and the number of existing patches, we need an automatic approach to identify recurring bugs as maintainers are busy with resolving newly reported bugs. The solution needs to be able to extract and summarize recurring bug patterns directly from existing patches. Furthermore, the patterns should be generic. Instead of matching a specific API call, e.g., a function name, it should match the code pattern, e.g., failing to release resources allocated by a specific function during errors. Traditional program analysis~\cite{zhou2022non,nielebock2020cooperative,yun2016apisan} methods typically depend on manual identification of bug patterns or extracted them by cross-checking numerous function calls. They also require dedicated, customized code analyzers for each specific bug pattern under review. Recent studies, such as Ullah et al.~\cite{ullah2024llms}, demonstrate that pure LLM-based methods still fall short in accurately detecting and analyzing bugs.

%
%

\PP{Our goal and approach}
We refer to these similar bugs as \textit{\textbf{Recurring Pattern Bugs (RPBs)}}. 
This term conveys that these bugs not only stem from recurring oversights but also have shared semantic characteristics, establishing a consistent pattern across multiple similar code implementations.
The goal of this project is to provide a systematic approach for identifying recurring pattern bugs. 
To this end, we propose \sys, a patch-centralized system that leverages a single patch as a clue to uncover all related bugs within a program.

Specifically, when a new patch is submitted by a reporter, \sys employs a static program analyzer to expand the patch's surrounding code. This step aims to enrich the patch's contextual information, as patches usually provide only a few lines surrounding the modifications, which is often insufficient for tasks such as patch summarization and bug identification.
\sys then leverages a large language model (LLM) to summarize the details of the patch into a precise coding rule while pinpointing the API functions or code pieces responsible for the bug.
After that, equipped with a static program analyzer, \sys further detects potential similar code implementations along with their associated context.
And in the final step, leveraging the capabilities of the LLM, \sys evaluates whether these identified instances are affected by the same issue fixed by the patch. 
This is achieved by crafting a well-structured prompt that integrates the preprocessed patch (optional), the derived security coding rule, and the target code instances that need to be analyzed.
As a result, \sys can autonomously adapt to bugs with recurring patterns without the need to develop specialized tools tailored to individual bug patterns.

To demonstrate the effectiveness of \sys, we evaluated it on a ground truth dataset of 80 recurring patterns and 850 corresponding recurring pattern bugs. 
On this dataset, \sys achieved 92.2\% precision and 79.1\% pairwise accuracy. 
We then applied 135 unique security coding rules to the Linux kernel, where \sys identified 22,568 potential violations, processing more than 117 million input tokens and producing more than 99 million output tokens. 
From a sample of 400 findings, 246 were confirmed as true violations likely to cause issues similar to the patched ones, including memory leaks, invalid pointer dereferences, and others.

	\begin{table*}[ht]
	\centering
        \scriptsize
        \resizebox{\textwidth}{!}{%
         \setlength{\fboxsep}{0.8pt}
\begin{tabular}{l|l|l|c|l|l|c}
\toprule
\textbf{API Name} &  \textbf{Bug Pattern}	& \textbf{Security Impact}	& \textbf{Commit ID} & \textbf{Module} & \textbf{Patch Time} & \textbf{Found by}  \\ 
\hline	

\multirow{3}{*}{\cc{iio_device_register_sysfs_group}} & \multirow{3}{*}{Missing release} & \multirow{3}{*}{Memory leak} & 95a0d596bbd0 & drivers/iio/industrialio-core.c & Dec 8 2023 & Static Analyzer \cite{liu2024detecting} \\ \cline{4-7}
& & & 86fdd15e10e4 & drivers/iio/industrialio-event.c & Nov 15 2022 & Fuzzer \\ \cline{4-7}
& & & 604faf9a2ecd & drivers/iio/industrialio-buffer.c & Oct 13 2021 & Fuzzer \\
\hline
\rowcolor{gray!20}
 &  &  & 884abe45a901 & drivers/net/..../en_accel/fs_tcp.c & Jun 28 2023 & Unknown \\ \cline{4-7}
\rowcolor{gray!20}
\cc{mlx5e_destroy_flow_table} & Missing nullification & Double free & e75efc6466ae & drivers/net/..../en/fs_tt_redirect.c & Nov 28 2023 & Static Analyzer\cite{liu2024detecting} \\ \cline{4-7}
\rowcolor{gray!20}
& & & 7a6eb072a954 & drivers/net/..../core/en_fs.c & Dec 28 2020 & Static Analyzer\cite{liu2021detecting} \\
\hline

\multirow{4}{*}{\cc{create_singlethread_workqueue}} & \multirow{4}{*}{Missing NULL check} & \multirow{4}{*}{NULL dereference} & 1fdeb8b9f29d& drivers/net/..../iwlegacy/3945-mac.c & Feb 8 2023 & Static Analyzer\cite{jiang2024app} \\ \cline{4-7}
& & & 41f563ab3c33 & drivers/parisc/led.c & Nov 17 2022 & Fuzzing \\ \cline{4-7}
& & & ba86af3733ae & drivers/net/...lan966x_ethtool.c & Nov 14 2022 & Fuzzing \\ \cline{4-7}
& & & a82268b30a8b & drivers/infiniband/hw/nes/nes_cm.c & Feb 17 2016 & Static Analyzer ~\cite{yun2016apisan} \\
\hline

\rowcolor{gray!20}
 & &  & de9b58400a3c & drivers/staging/..../ioctl_linux.c & Jul 12 2018 & Compile Warning \\ \cline{4-7}
 \rowcolor{gray!20}
\cc{strncpy} & Using unsafe API & Overflow & 81b9de43599c & drivers/media/media-device.c & Jan 8 2018 & Compile Warning \\ \cline{4-7}
\rowcolor{gray!20}
& & & b3f8ab4b7953 & fs/9p/vfs_inode.c & Jul 16 2013  & Unknown \\ \cline{4-7}
\hline


\bottomrule
\end{tabular}

        }
        \caption{Examples of Recurring Pattern Bugs in the Linux kernel across four APIs. Similar issues needed to be fixed multiple times across the kernel.}
        \label{t:hpbStudy}
    \end{table*}
    
To summarize, we made following contributions:
\begin{itemize}
    \item \textbf{An LLM assisted program analysis framework for bug identification.}
    We introduce a framework, \sys, that combines lightweight static analysis with LLM reasoning to identify recurring pattern bugs (\hpb) from a single exemplar patch. 
    The system can discover previously unseen, similar bugs without building specialized analyzers for each bug pattern. 
    To our knowledge, this is the first work to apply LLMs to large scale code analysis for bug detection. 
    In our large scale analysis, \sys analyzed more than 148K code pieces, processing 117 million input tokens and producing 99 million output tokens. 
    These artifacts provide a useful basis for future studies of LLM based code reasoning.

    \item \textbf{A ground truth dataset of recurring pattern bugs.}
    We manually constructed a ground truth dataset containing 850 security patches grouped into 80 recurring patterns, derived from more than 1,900 patches reported across 23 state-of-the-art works. 
    This dataset can support future research on LLM based code understanding and \hpb detection.

    \item \textbf{A systematic evaluation of LLMs for code analysis and bug finding.}
    We evaluated six LLMs under six prompt configurations on the ground truth dataset and identified factors that affect accuracy and cost, such as prompt design and information sources. 
    These results offer practical guidance for applying LLMs to program analysis.

    \item \textbf{New recurring pattern bugs identified in the Linux kernel.}
    \sys uncovered more than 22K potential security coding rule violations in the latest Linux kernel. 
    From a sample of 400 cases, 246 were manually confirmed as valid. 
    These cases can cause a range of security impacts, including invalid pointer dereferences, resource leaks, and type errors. 
    In collaboration with the Linux maintainers and security researchers, we are actively prioritizing, validating, reporting, and fixing these findings at scale.

\end{itemize}

\section{Background}
\label{s:background}

We first define the key terms used throughout the paper. 
We then characterize recurring pattern bugs (\hpb), examining their origins, prevalence, diversity, and security impact.

\subsection{Terminology}
\label{s:term}
This section defines terms used throughout this work:

\PP{Code Piece} A continuous segment of source code such as a single macro, several basic blocks, or an entire function.

\PP{Recurring Pattern} A repeated usage pattern of an API or code piece.

\PP{Recurring Pattern Bugs (RPBs)}
Recurring and similar errors that arise from misusing the same API, code piece, or code pattern.

\PP{Security Coding Rule}
A concise statement specifying correct usage of an API or code piece; violating this rule may introduce security vulnerabilities.

\PP{Seed Patch}
A representative patch that fixes a single instance of an RPB. It can be used as a concrete example to identify \hpb or to generate the corresponding security coding rule.


\subsection{Characteristics of Recurring Pattern Bugs}
\label{bugstudy}

To understand why these bugs occur, we analyzed the Linux kernel as it is a large, well-maintained, active open-source project with an extensive patch history available for mining. We manually reviewed patches for the Linux kernel and identified several APIs with multiple patches targeting similar bugs. These bugs vary in security impact, but are all instances of \hpb. Examples and relevant information about these bugs are presented in \autoref{t:hpbStudy}. Upon review, we find that \hpb occur due to: 
\begin{enumerate}
    \item \textbf{Repeated mistakes:} Developers can misuse API functions or reuse buggy code pieces due to unfamiliarity with the code. For example, one common mistake~\cite{he2023one} is incorrectly incrementing or decrementing a reference count due to a misunderstanding of the semantics of wrapper functions around refcount operations.
    
    \item \textbf{Omission of security safeguards:} Some API functions or code pieces can fail, so they need to be accompanied by error handling code, usually by the caller, that safeguards against potential security issues. Many \hpb occur because these safeguards, such as return value checks or pointer nullification, are missing.

    A previous patch‑analysis study~\cite{wu2020precisely} showed that security vulnerabilities often arise from the interplay of multiple bug conditions. For instance, releasing a pointer and later dereferencing it can create a use‑after‑free bug. The original code may correctly implement an API call like \cc{kfree(P)} on the pointer \cc{P}; however, a code change may introduce a dereference of \cc{P} after \cc{kfree(P)}. This could occur in the same function where \cc{kfree(P)} is called, but could also occur elsewhere, a different function or even another thread, resulting in a hidden use‑after‑free bug. When such safeguards are missed, \hpb  occur.
\end{enumerate}
%


%

\begin{figure}[h]
    \centering
\includegraphics[width=.98\columnwidth]{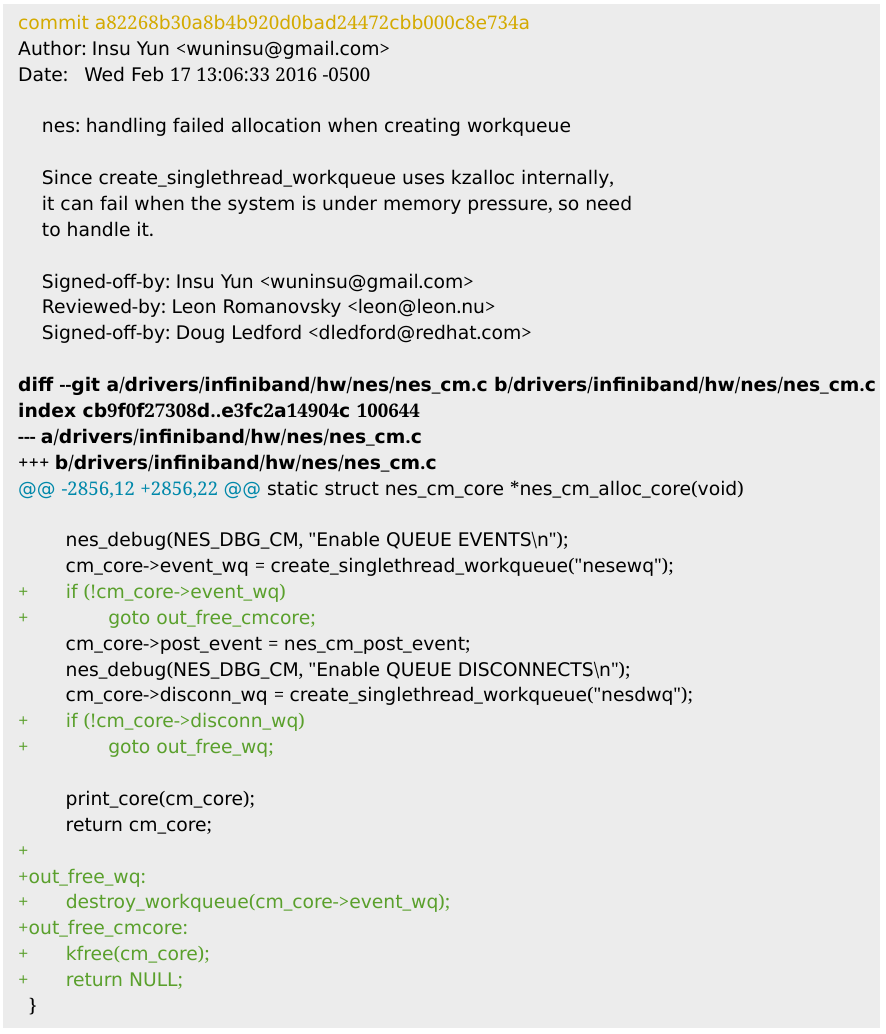}
    \caption{The \cc{create_singlethread_workqueue} macro can return \cc{NULL} on failure. This patch adds a simple check to \cc{nes_cm.c}. It is invoked more than 197 times in other parts of the Linux kernel, but not every invocation checks the returned pointer value. This creates an RPB.}
    \label{fig:null_example}
\end{figure}

\PP{Ease of Occurrence}
Whether a single bug eventually becomes part of an RPB depends on three factors: (1) the total number of times the misused API or code piece is invoked, (2) how easily it can be misused, and (3) how consistently developers apply its preventive safeguards if any exist.
Consider the macro \cc{create_singlethread_workqueue}, which has a patch shown in \autoref{fig:null_example}. 
This macro may return \cc{NULL} on failure. Therefore, the patch adds the missing safeguard: check if the returned pointer is \cc{NULL} and perform any necessary cleanup. However, this macro is invoked more than 197 times in the Linux kernel and not every invocation is properly safeguarded. Our review of the Linux kernel’s patch history found 16 separate fixes, four of which are highlighted in \autoref{t:hpbStudy}, for call sites that omitted this safeguard, leading to NULL pointer dereferences.

In some cases, certain APIs are more error prone, 
 which makes \hpb for these APIs more common. He et al.~\cite{he2023one}, for example, report that the power management function \cc{pm_runtime_get_sync} introduced more than ninety bugs despite being called only a few hundred times in the kernel.
This function unconditionally increments its internal reference count, deviating from the usual ``increment on success'' pattern of resource acquisition APIs. This deviation often causes developers to fail to decrease the corresponding refcount.

\PP{Time and Security Impact of RPBs}
As shown in \autoref{t:hpbStudy}, \hpb can manifest in the same module or span multiple modules and may be the result of contributions by a single developer or multiple developers over time. Modularized maintenance makes it difficult to identify and fix \hpb due to lack of awareness of related patches and complex dependencies. Furthermore, although large projects regularly enforce coding standards, developer diversity inevitably causes some divergence across the code base, thus increasing the odds of an RPB. Once introduced, either due to a new invocation or a change in API, it may remain unaddressed for years. For example, the macro \cc{create_singlethread_workqueue} allocates a work queue and will return \cc{NULL} on failure, and its return value must be checked to avoid a NULL‑dereference. According to the patch history, we found four other patches across a span of seven years (from 2016 through 2023) across multiple drivers fixing unprotected use of this macro. \emph{Limitations in existing detection methods often only reveal a few isolated instances of an RPB.} Fuzzing, which was used for four of the 13 patches we studied, only fixed the triggered issue, rather than all issues across the kernel. Moreover, \hpb give rise to a wide spectrum of security consequences.
As \autoref{t:hpbStudy} illustrates, they can manifest as memory leaks, NULL‑pointer dereferences, reference‑count leaks, use‑after‑free errors, and buffer overflows. 
%


\ignore{
\subsection{Limitation of existing approach in finding and fixing the Recurring Pattern Bugs}

\PP{Dynamic analysis focus on the memory impacts of bugs} 
dynamic analysis focus on the impact of bugs: out of bounds buffer access; memory uses after free; NULL dereference; 
Limited coverage scope; 
Limited coverage of types: mainly memory related ones.

\PP{Static analysis requires human specified bug pattern}
Some features of recurring bugs may be captured, but not representative enough;
Need to manually implement the analyzer for find each pattern of bug;

\PP{Single patch for single bug strategy}
Independent reports for similar Recurring bugs
Patch: handle the vulnerable conditions introduced by the API

also check: d640627663bfe7d8963c7615316d7d4ef60f3b0b

how to know if this issue is HPB??? is hard to know that.

pattern matching?: check vim net/sctp/socket.c +2214
}

\ignore{
\yue{TODO: }
\subsection{Large language model for code understanding and bug detection}

\QW{Differentiate this subsection with the related work section}

\PP{LLM for code understanding.} 
LLMs are increasingly utilized for code understanding tasks across various stages of software development~\cite{nam2024using,haroon2025accurately}.
These models are typically pre-trained on large corpora of natural language and source code, enabling them to learn both syntactic and semantic patterns in programming languages~\cite{feng2020codebert,ma2024unveiling,chen2021evaluating}.
For instance, OpenAI Codex~\cite{OpenAICodex} was trained on over 100 million public GitHub repositories, encompassing dozens of terabytes of source code across multiple languages.
When adapted to code understanding, LLMs can reason over function-level logic~\cite{ma2023lms}, infer data and control dependencies~\cite{wang2024llmdfa}, reasoning program invariants~\cite{pei2023can}, and detect anomalous or non-idiomatic usage patterns~\cite{ahmad2023flag} that may indicate defects.
Unlike traditional program analysis techniques—which rely on formal semantics and handcrafted rules to provide precise and exhaustive analysis—LLMs operate on learned representations and are capable of generalizing across diverse programming contexts, offering a more intuitive, human-like understanding of code~\cite{yang2025evaluating}.
Moreover, they can be quickly adapted to new tasks through few-shot or in-context learning, requiring minimal engineering effort compared to traditional static analyzers, which depend on manual rule configuration.
As a result, LLMs are emerging as a powerful and complementary tool for program comprehension, particularly in scenarios where rule-based systems fall short or are difficult to scale.

\PP{LLM for bug identification}
Building on their ability to understand and reason about source code, LLMs have recently been applied to automated bug identification tasks~\cite{}. These applications span a variety of domains, including detecting use-before-initialization (UBI) bugs in the Linux kernel~\cite{li2024enhancing}, identifying defects in graph database engines~\cite{wu2024effective}, and uncovering analogous bugs in deep learning frameworks~\cite{guan2025crossprobe}.
Because LLMs are trained on vast amounts of code—including both correct and buggy examples—they are capable of learning patterns commonly associated with programming errors~\cite{}.  
These errors may involve unsafe memory access~\cite{}, misuse of APIs~\cite{}, or violations of standard coding practices~\cite{}. By leveraging these learned patterns, LLMs can flag suspicious or incorrect code in a way that resembles human reviewers who draw on experience with similar bugs.
LLMs have also been shown to support different modes of bug detection. They can classify whether a given function is buggy~\cite{}, point out specific lines that may be problematic~\cite{}, or explain in natural language why a bug might exist~\cite{}. In some cases, they can even suggest patches or rewrites to fix the bug~\cite{}. These emerging capabilities have led to growing interest in using LLMs for tasks such as vulnerability triage, patch review, and automated code auditing.

\PP{Limitations} 
Despite their promising capabilities, LLMs are not without limitations.
First, they are prone to hallucinations—generating plausible but incorrect outputs—which can lead to misidentified bugs or misleading explanations~\cite{ji2023survey,huang2025survey}. Their responses may also vary across runs and are constrained by limited context windows, reducing reliability in large or multi-file code bases~\cite{}.
In addition, LLMs are resource-intensive: models like CodeLLaMA-34B require significant GPU memory and energy~\cite{code-llama-blog}, while commercial APIs (e.g., GPT-4) can cost \$0.03–\$0.06 per 1,000 tokens~\cite{gpt-cost}, limiting scalability. 
%
What's more, latency presents a practical challenge. GPT-4 exhibits a Time to First Token (TTFT) of approximately 0.6 seconds, while other models like Claude and Gemini 2.9B can take between 1 and 14 seconds~\cite{llm-performance}, rendering them unsuitable for real-time development workflows that require immediate feedback.

\yue{For the following points, should we include them in the Discussion or Measurement section, since these insights are derived from our experiments?} \QW{Oh! don't need to worry that, those are just some random thoughts I had previously. you can write with you style :)}
}
\section{Workflow of \sys}
\label{s:overview}

\begin{figure*}[htbp]
\centerline{\includegraphics[width=\linewidth]{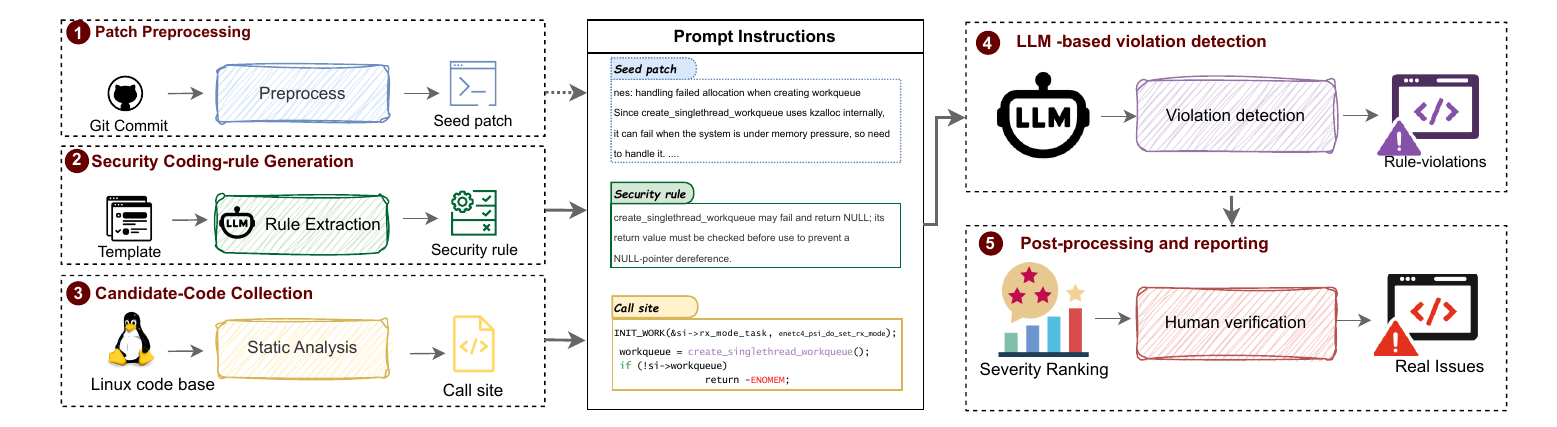}}
\caption{Overview of \sys.}
\label{fig:overview}
\end{figure*}

\autoref{fig:overview} illustrates the end‑to‑end workflow of \sys.
First, \sys is provided with a Git commit corresponding to a bug fix. Information not pertinent to the fix is removed (e.g., diff headers, URLs, etc.) and the complete function containing the fix is identified. This forms a seed patch, contains the  ``cleaned'' patch information and the surrounding code context. Second, the seed patch is sent to an LLM via a prompt template (see \autoref{t:prompt_rule}) that instructs the LLM to propose a concise rule specifying the correct usage pattern of the misused API or code piece. 
Third, a lightweight static analyzer scans the entire code base, 
collects all call sites of the API or code segment associated with the recurring pattern, and marks them as potential \hpb candidates.
Fourth, for each candidate, \sys combines code context of the call site with the extracted security rule and the seed patch to create an LLM prompt. The prompt instructs the language model to judge whether the call site violates the security rule (see \autoref{t:prompts} for the full template).
Finally, \sys aggregates the model’s responses across all candidates, performs post-processing to validate, and presents suspected violations to human developer for manual review. 
All findings that persist after human verification are sent for patching.

For example, consider commit \cc{a82268b30a8b} in \autoref{fig:null_example}. In step 1, \sys forms a seed patch by removing the irrelevant metadata in the patch information and extracting \cc{nes_cm_alloc_core} function which contains the misuse of the \cc{create_singlethread_workqueue} macro. 
In step 2, the seed patch is provided to the LLM extractor, which proposes the following security rule: \textit{``\cc{create_singlethread_workqueue} may fail and return \cc{NULL}; its return value must be checked before use to prevent a NULL‑pointer dereference.''}. 
In step 3, the static analyzer scans for call sites of \cc{create_singlethread_workqueue} and finds 197 code segments. These code segments are processed by the LLM in step 4 via the violation instruction prompt and \sys flags 10 new cases as potential violations. 
Manual review identified four true violations, 
three of which are bugs similar to the one fixed by the seed patch.

\section{Methodology}
\label{s:desig}

\label{s:rchallenge}

Ideally, given a single bug, the tool should surface other \hpb in analyzed code bases. To remain general across recurring patterns, we avoid relying solely on traditional static or dynamic analyses. 
Dynamic methods, such as fuzzing, explore executions randomly and primarily reveal a narrow set of memory errors under specific sanitizers.
They are poorly suited to catching bugs that arise only under specific call sequences or state conditions including \hpb.
Traditional static-analysis approaches face a different challenge.
Each bug pattern typically demands a specialized analyzer that models its control and data flow semantics, followed by iterative constraint tuning. 
Engineering a new checker for every patched pattern is impractical at scale.

Therefore, we aim to meet three key requirements:
(1) providing a systematic method to construct prompts and contexts that leverage the strengths of LLMs when analyzing code and detecting diverse types of \hpb, while minimizing manual and engineering effort,
(2) ensuring the solution is both cost-efficient and effective, and
(3) achieving accuracy that is comparable to state-of-the-art bug finding tools.

\PP{LLM-based semantic analyzer for flexible code understanding}
The above considerations motivate the code analysis module of \sys to adopt an LLM-based approach. Given the cross-domain language diversity in patches, natural language and code, LLMs are well suited to extracting the pattern from a given patch and performing pattern matching. The LLM ecosystem is also diverse and constantly evolving, which provides us with the ability to swap models based on user requirements and ensure state-of-the-art performance. However, we do not rely entirely on the LLM for all aspects of the vulnerability detection. \sys is a rule-centric, patch-seeded, LLM-driven workflow, augmented with lightweight static enumerations to efficiently discover similar bugs. 
Unlike traditional program analysis-based research that focused on developing highly sophisticated analysis techniques, the main contributions of this work are LLM-assisted analysis framework and systematic study with real cases demonstrating how LLMs can be leveraged for code analysis and bug discovery.

\PP{Improved patch information}
Patches involve a mix of natural language information and code information. The natural language information contains user written details of the issue, and sometimes, the fix, but can also include extraneous information such as email headers, timestamps, discovery method, URLs, etc., which can distract an LLM and increase inference costs. \sys applies deterministic preprocessing that extracts only the user submitted issue description from the natural language information to minimize inference costs and improve detection performance.

The code information also requires preprocessing. Standard diffs expose only the edited lines with minimal code context. It can be difficult for an LLM to infer the reason for the edits or generalize the misuse pattern with such little information. 
This is particularly challenging when the root cause could exist dozens of lines away from the edits, e.g., a memory leak in subsequent error paths of an allocation, and be outside the context.
\sys expands the code context to encompass the entire function, showing the edited lines within that function-wide perspective. This enriched code context preserves the control flow and data dependencies that may be essential for bug detection. Cleaning the patch metadata and expanding the code context creates a seed patch that is processed by the detection LLM.

\PP{Lightweight program analysis as a function level context builder}
The main challenge is finding \hpb in an existing code base, especially for large projects. Model context limits and high token costs make it inefficient to supply entire files, not to mention a full code base, to an LLM and ask it to search chunks at a time. This approach also increases the chances of false positives and false negatives given the nondeterministic and unreliable nature of LLMs, e.g., hallucinations. Previous studies of security fixes~\cite{li2017large, wu2020precisely} show that most patches and their relevant context span only dozens of lines, suggesting that short, focused snippets are often sufficient. Therefore, \sys uses a lightweight static analyzer to preprocess the code base and enumerate all call sites that match the target API or code piece potentially associated with an RPB. As with patches, for each call site, \sys extracts the entire caller function.
%

We note that an additional benefit of using functions for the code context  is improved extensibility. Although intra-procedural analysis is sufficient to detect many \hpb, more complex bugs may require inter-procedural reasoning. With function-level context in place, in the future, extending \sys using call graphs to support inter-procedural analysis becomes straightforward.

\PP{Bug detection prompt construction}
Without explicit information about a bug pattern, LLMs cannot reliably detect security bugs, often performing even worse than random guessing (see \autoref{ss:accuracy}). To ground the LLM and focus on identifying a specific bug pattern, we evaluate three approaches for processing a seed patch and transforming it into an RPB identification prompt. In the patch-based approach, we provide the seed patch and the target code snippet to the LLM and ask it if the bug fixed in the patch is also present in the target code snippet. In the rule-based approach, we apply an additional processing step that converts the seed patch into a security coding rule. Then, it instructs the detection LLM to assess if the target code snippet violates the extracted rule. In the hybrid approach, we provide both the seed patch and the extracted security rule and ask the detection LLM to assess if the target code snippet violates the extracted rule. As we show in \autoref{ss:accuracy}, the rule-based approach empirically yields the best trade-off: higher accuracy with lower cost compared to the other design choices.

%
%

%

\PP{Security rule generation}
A Security coding rule used for RPB detection must precisely capture both the recurring pattern bug and the underlying root cause of the bug. We explore two different approaches for obtaining an RPB security coding rule: human generation and seed patch generation. When reviewing a patch, a maintainer can manually define a rule and provide it to \sys to find other \hpb instances. This approach is useful for novel \hpb, but human generated rules can vary in expressiveness and details. For example, one expert might write ``free buffers allocated by \cc{framebuffer_alloc} using \cc{framebuffer_release},'' while another writes ``use \cc{edac_mc_free} to release buffers from \cc{edac_mc_alloc} in the error path.''  Both are rules on how to fix a specific memory leak issue, but can unintentionally add or remove constraints to the detection pattern.

Our seed patch-based rules leverage an LLM to automate rule generation from a seed patch. As generation is automatic, we need to balance rule granularity. Rules that are too abstract may fail to constrain bug patterns and over report; rules that are too detailed may overfit to data flow or implementation specifics and do not generalize well. Based on our study of more than 1,900 security bug patches, we define a high-quality rule as the one that specifies the root cause (the API or code piece whose misuse causes the bug), the required handling action to prevent the bug, the expected impact if the bug is triggered, and any scope constraints on applicability, such as restriction to error-handling paths or particular classes of functions. We deploy a standardized, concise set of rule templates and a prompting procedure that yields outputs informative enough to guide detection, yet compact enough to generalize and assist maintainers. See \autoref{t:prompt_rule} for the prompt to generate a security coding rule from an RPB patch.

\section{Evaluation}
\label{s:eval}

In this section, we evaluate \sys on an expert-annotated ground truth dataset. 
All bugs in the dataset were identified by security studies published at top-tier conferences. 
We assessed \sys with both open-source and closed-source models, reporting accuracy, cost, and ablation results.
\autoref{ss:env} describes the environment setup,
\autoref{ss:gtcollection} details the ground truth collection, and
\autoref{ss:accuracy} presents accuracy and ablation results, quantifying the contribution of each system component. 
Finally, \autoref{ss:cost} compares cost and performance across different models.

\subsection{Environment setup}
\label{ss:env}

\PP{Model Selection} 
In this evaluation, we tested six LLMs with \sys using identical hyperparameters. 
Specifically, we set temperature, top-$p$, and $n$ to 1, following the default configuration of the OpenAI chat completion API~\cite{hyperparameters,hyperparametersllama,hyperparametersclaude}.
For state-of-the-art open-source models, we used models from the Llama family. For state-of-the-art closed-source models, we used models from the GPT and Claude families.
Specifically, we evaluated Llama-3.3-8b and gpt-4.1-nano as lightweight, low-cost models; 
Llama-3.3-70b and Llama‑4-17B-128E as mid-sized models that balance cost and accuracy; 
and o4-mini and Claude Sonnet 3.7 as higher-cost frontier models.

\begin{table}
\footnotesize
        \resizebox{\columnwidth}{!}{%
        
\begin{tabular}{l|c|c|c|c}
\toprule
\multirow{2}{*}{\textbf{Setup}} & \multicolumn{3}{c|}{\textbf{Input}} & \textbf{Strategy} \\\cline{2-5}
 & \textbf{Code Snippet} & \textbf{Patch} & \textbf{Coding Rule} & \textbf{CoT} \\\hline
\rowcolor{gray!20}
\textbf{Basic} & \cmark & & & \cmark \\
\textbf{Patch} & \cmark & \cmark & & \cmark \\
\rowcolor{gray!20}
\textbf{HuRule}& \cmark & & Human & \cmark \\
\textbf{Rule}  & \cmark & & System & \cmark \\
\rowcolor{gray!20}
\textbf{Rule w/o CoT} & \cmark & & System & \\
\textbf{Rule + Patch} & \cmark & \cmark & System & \cmark \\
\bottomrule
\end{tabular}

        }
\caption{Prompt Configurations}
\label{tab:prompt_configuration}
\end{table}

\PP{Prompt Configurations}  We evaluated the following six prompt configurations to isolate and study the influence of the subcomponents of \sys. The differences and detailed prompts are provided in \autoref{tab:prompt_configuration} and \autoref{t:prompts}.

\begin{itemize}
    \item \textit{Basic configuration.} The model receives only the code snippet and is asked to determine whether it contains bugs, without any further detailed bug information, security rules, or examples. 
    
    \item \textit{Patch configuration.} The model receives the code snippet together with a seed patch from the same class and is asked to determine whether the code snippet contains bugs similar to the one fixed by that patch.
    
    \item \textit{HuRule (Human generated rule) configuration.} A human security expert first defines a security coding rule for the RPB class. The model uses that rule to decide whether the provided code snippet has violated the security rule, thus representing a bug.
    
    \item \textit{Rule configuration.} Instead of the human generated rule, each model receives a formatted security coding rule, automatically extracted by \sys from the patch. The model uses that rule to decide whether the provided code snippet has violated the security rule, thus representing a bug.
     
    \item \textit{Rule w/o CoT configuration.} Building on the rule configuration, we removed the chain-of-thought guidance sentence: ``Analyze the code line by line and show the reasoning steps.'' This variant is used to compare against the original rule configuration in order to study the impact of chain-of-thought prompting strategies on code understanding and bug detection task.

    \item \textit{Rule+Patch configuration.} The model receives both the \sys generated rule and the seed patch. The model uses both items to decide whether the provided code snippet has violated the security rule, thus representing a bug.
    
\end{itemize}

\PP{Evaluation Metrics}
For each prompt configuration and model, we report the precision, recall, and pairwise accuracy to quantify performance.
Precision ($\mathcal{P}$) is the fraction of reported positives (i.e., flagged as an RPB) that are correct.
Recall ($\mathcal{R}$) is the fraction of true bugs that are detected. 
Higher recall reflects better coverage and robustness.
Pairwise accuracy ($\mathcal{PA}$) is a strict measure computed per patched and unpatched pair: the model earns credit only if it labels the patched snippet as safe and the unpatched snippet as buggy. 
Any other outcome counts the pair as incorrect.
Thus, a random guess yields $\mathcal{PA}=25\%$. 
The formulas for these metrics are shown below.

\begin{equation}
\mathcal{P} = \frac{TP}{TP + FP}
\end{equation}

\begin{equation}
\mathcal{R} = \frac{TP}{TP + FN}
\end{equation}

\begin{equation}
\mathcal{PA} = \frac{\# \text{ of Correctly Labeled Pairs}}{\# \text{ of Total Pairs}}
\end{equation}

\PP{Human Effort} 
In this project, we engaged six security researchers, each holding a Ph.D. in a relevant security field and possessing between six and thirty years of experience in the security domain. 
These experts performed all manual assessments, including ground-truth collection and labeling, analysis of false positives and false negatives, and verification of detection results.


\subsection{Ground truth Collection}
\label{ss:gtcollection}

	\begin{table*}[t]
	\centering
        \footnotesize

            \resizebox{\textwidth}{!}{%

\newcolumntype{P}[1]{>{\centering\arraybackslash}p{#1}}

\begin{tabular}{l|ccc|ccc|ccc|ccc|ccc|ccc}
\toprule
\textbf{Setup} &
\multicolumn{3}{c|}{\textbf{gpt-4.1-nano}} &
\multicolumn{3}{c|}{\textbf{Llama-3.3-8b}} &
\multicolumn{3}{c|}{\textbf{Llama-3.3-70b}} &
\multicolumn{3}{c|}{\textbf{Llama-4-17b-128E}} &
\multicolumn{3}{c|}{\textbf{Claude Sonnet 3.7}} &
\multicolumn{3}{c}{\textbf{o4-mini}} \\
\midrule
\textbf{Metrics} & $\mathcal{P}$ & $\mathcal{R}$ & $\mathcal{PA}$ 
                 & $\mathcal{P}$ & $\mathcal{R}$ & $\mathcal{PA}$ 
                 & $\mathcal{P}$ & $\mathcal{R}$ & $\mathcal{PA}$ 
                 & $\mathcal{P}$ & $\mathcal{R}$ & $\mathcal{PA}$ 
                 & $\mathcal{P}$ & $\mathcal{R}$ & $\mathcal{PA}$ 
                 & $\mathcal{P}$ & $\mathcal{R}$ & $\mathcal{PA}$ \\
\midrule

\rowcolor{gray!20}
\textbf{Basic} & 54.3\% & 78.1\% & 20.0\%  & 49.4\% & 30.5\% & 16.8\% & 53.8\% & 57.7\% & 22.0\% & 53.5\% & 83.4\% & 17.3\%&         58.2\% & 81.8\% & 29.1\% & 61.3\% & 85.1\% & 37.3\% \\ \hline

\textbf{Patch}  & 69.5\% & 72.2\% & 50.9\% & 71.4\% & 55.1\% & 39.0\% & {90.7\%} & 67.4\% & 62.6\% & 88.0\% & 72.5\% & 63.9\% &       \textbf{91.2\%} & 79.1\% & 71.9\% & 89.6\% & 87.4\% & 77.8\% \\ \hline

\rowcolor{gray!20}
\textbf{HuRule}& 80.4\% & \textbf{84.5\%} & 67.7\%  & \textbf{83.4}\% & \textbf{67.8}\% & \textbf{57.9\%} & 87.2\% & \textbf{78.8\%} & 69.4\% & 88.8\% & 81.4\% & 72.8\% &           88.5\% & 87.7\% & 76.7\% & 88.7\% & 90.9\% & 79.9\% \\ \hline

\textbf{Rule} & \textbf{82.2\%} & \textbf{85.3\%} & \textbf{70.1\%} & 79.1\% & 64.9\% & 52.5\% & {90.6\%} & \textbf{78.3\%} & \textbf{71.4\%} & \textbf{92.2\%} & \textbf{85.8\%} & \textbf{79.1\%}  &            \textbf{91.5\%} & \textbf{89.0\%} & \textbf{80.8\%} & \textbf{90.3\%} & \textbf{92.9\%} & \textbf{83.1\%} \\ \hline

\rowcolor{gray!20}
\textbf{Rule w/o CoT} & 55.7\% & 52.2\% & 23.7\% & 73.6\% & 19.9\% & 14.4\% & \textbf{94.9\%} & 7.3\% & 7.3\% & \textbf{92.9\%} & \textbf{85.3\%} & \textbf{79.5\%}  &          88.1\% & 86.0\% & 74.6\% & \textbf{90.5\%} & \textbf{93.1\%} & \textbf{83.5\%} \\ \hline

\textbf{Rule+Patch} & 80.8\% & 68.4\% & 57.5\% & 80.9\% & 64.8\% & 55.0\% & {90.4\%} & \textbf{78.7\%} & \textbf{71.3\%} & {91.8\%} & 83.1\% & 77.0\% &        \textbf{91.4\%} & \textbf{88.8\%} & \textbf{80.9\%} & \textbf{91.3\%} & \textbf{92.7\%} & \textbf{84.1\%} \\ \hline

\bottomrule
\end{tabular}
        }

        \caption{Accuracy of LLMs under different setups ($\mathcal{P}$=Precision, $\mathcal{R}$=Recall, $\mathcal{PA}$=Pairwise Accuracy). Best results are bolded. Differences <1\% are also bolded and treated as similar performance due to non-determinism.} 
        \label{t:ablationsT}
    \end{table*}

To evaluate \sys on real-world \hpb in the Linux kernel, we built a ground truth dataset of \hpb. 
We surveyed 23 security studies published at top tier conferences that analyze the Linux kernel and report vulnerabilities~\cite{lu2019detecting, HERO,pakki2020exaggerated,emamdoost2021detecting,liu2021detecting,zhou2022non,liu2024detecting,wang2018check,wang2021mlee,yun2016apisan,min2015cross,jeong2019razzer,zhai2022progressive,lyu2022goshawk,tan2021detecting,suzuki2024balancing,dossche2024inference,lin2023detecting,lin2025uncovering,defreez2018path,kim2020hfl,xu2018precise,bai2019effective}.
To collect the bugs found by these studies, we extracted author names and email addresses from the papers and matched them with the accepted commits in the Linux kernel. 
This process identified 29 authors who contributed at least one accepted patch and yielded 1,910 patches in total.

We manually reviewed all 1,910 commits to identify patches for \hpb. 
To construct a ground truth dataset, a recurring pattern is defined as at least two patches that address the same misuse of a given API or code piece. We manually annotated each identified pattern with a security coding rule. This resulted in 80 recurring patterns and associated rules, comprising a total of 850 patches.
We also identified 39 independent patches that fix specific API or small code fragment misuse, but have no similar cases among the 1,910 patches.
We annotated rules for these cases as well.
Although they can still support large‑scale \hpb detection, we exclude them from the ground-truth evaluation, which requires at least two patches per pattern.
The remaining 1,021 patches fall outside the scope of this work, for example those involving concurrency design problems, complex inter-procedural dependencies, as well as those involving code functionality changes.

As a result, the \textbf{850} patches grouped into 
\textbf{80} recurring patterns constitute our ground truth dataset. 
From each pattern, we randomly selected one patch as a designated seed patch, which is excluded from evaluation. The remaining patches in the pattern are used as evaluation targets.
For every target, we extracted both the unpatched code (i.e., positive case, rule-violating) and the patched code (i.e., negative case, rule-respecting). Thus, for each prompt configuration, we evaluated 1,540 snippets: $2\times(850 \text{ patches} - 80 \text{ seed patches})$.
%

\subsection{Performance and Ablation study}
\label{ss:accuracy}

\autoref{t:ablationsT} summarizes the performance of the six LLMs on our ground truth dataset under the six prompt configurations (see \autoref{appendix} for full prompt details).
Each setup defines a distinct prompting scenario for the LLM to detect bugs in a code snippet. 

\PP{LLMs fail to identify bugs without context (Basic Setup)} 
With the Basic configuration, most models performed at or below the random guessing baseline of 25\% $\mathcal{PA}$.
Llama‑4, Claude Sonnet 3.7, and o4‑mini in particular tend to label code as buggy, boosting recall but generating more false positives causing to waste human effort on verification in real-world bug finding.
These results confirm prior findings that, without targeted guidance, LLMs struggle to accurately detect real‑world bugs~\cite{ullah2024llms}.
By contrast, the other five configurations, providing a patch, a human‑crafted rule, an auto‑formatted rule, or both patch and rule, substantially improve performance by enabling the model to focus on the relevant bug pattern.

\PP{LLMs can detect RPBs with moderate accuracy but at higher cost using seed patches (Patch Setup)}
The Patch configuration boosts performance by 22\% to 47\% of $\mathcal{PA}$ over the Basic setup.
This suggests that providing a seed patch as a detailed example of \hpb in the same class helps LLMs find other \hpb. 
However, this capability depends strongly on model capacity, which means that not every model can correctly interpret and apply the information in the example. 
For instance, smaller models such as Llama-3.3-8b and gpt-4.1-nano still struggle to use the example to detect \hpb accurately. 
By contrast, stronger frontier models like Claude Sonnet 3.7 and o4-mini can identify \hpb from a seed patch with substantially better precision and $\mathcal{PA}$.
This indicates that giving a detailed example of one of the \hpb in the same class can assist LLM in finding similar bugs.

Besides heavily relying on frontier models, the Patch configuration has two additional disadvantages. 
First, the Patch configuration is considerably more expensive than the Rule configuration.
Because the patch plus the context are far larger than security coding rules, and require on average 13.2K extra input tokens per case (see \autoref{t:cost}), incurring significantly higher inference costs.
Second, compared with the Rule configuration, the Patch configuration yields lower precision. 
The performance drop in relative accuracy stems from inconsistent patch quality, complexity of the patch, and the abundance of extra context (see Section \autoref{s:rchallenge}), which can distract the model.

\PP{Providing a formatted security coding rule enables precise RPB detection (Rule Setup)}
The Rule configuration yields the best $\mathcal{PA}$ and a high precision across most LLMs. 
Notably, Llama‑4 achieves 92.2\% precision and 79.1\% $\mathcal{PA}$ under this configuration; given its strong accuracy, throughput, and cost profile, we select Llama‑4 as our preferred model for large‑scale \hpb detection (see Section \autoref{s:study}). 
Therefore, for this project we select the formatted rule configuration as our primary choice, because it is lower cost, delivers higher accuracy across models, and supports easy automatic generation.

\PP{Using human generated rule or using patch with rule may not add additional performance to the model (HuRule and Rule+Patch Setups)} 
In addition to testing the formatted rule configuration alone, we evaluated two alternatives, the formatted rule combined with a seed patch, and the human expert-generated rule. 
Compared to formatted rules, both seed patches and human written rules are subject to natural language variability, as discussed in Section \autoref{s:rchallenge}. 
Consequently, their quality and specificity can vary.
Frontier models, for example o4-mini, show modest improvements when supplied with both the patch and the rule. 
This likely stems from the concrete example in the patch plus an extracted abstract rule better guiding high capacity models. 
However, the token count more than doubles, which greatly increases the inference cost, so whether the small accuracy gain justifies that cost remains an open question.
Interestingly, Llama-3.3-8b achieves its best results with the human-generated rule compared with other configurations.
Human-written rules use plain, conversational language similar to the training data for the model, which we think makes these rules easier for this smaller model to interpret.
Even so, its precision and pairwise accuracy remain far below those of larger models and are not yet adequate for real-world deployment.

\PP{Small models need chain of thought, limited effect on larger models (Rule w/o CoT Setup)}
We evaluated the formatted rule setting after removing a single instruction for chain of thought (CoT), namely ``Analyze the code line by line and show the reasoning steps.''
Without chain of thought, pairwise accuracy for Llama-3.3-8b, Llama-3.3-70B, and gpt-4.1-nano approaches or falls below the 25 percent random baseline. 
Examining the output of these models shows that these models often return simple ``Yes'' or ``No'' with little explanation, or fabricate a short rationale that explains the chosen label. 
Within the Llama-3 family, the models frequently answered ``No'', which raised precision but sharply reduced recall and pairwise accuracy. 
gpt-4.1-nano yielded roughly balanced yet largely random ``Yes'' and ``No'' decisions, producing a pairwise accuracy near 25\%. 

In contrast, larger frontier models are more resilient. 
Removing chain of thought has only a modest effect on accuracy, sometimes even a slight improvement, and it also reduces output length. 
Notably, for o4-mini without explicit chain of thought prompting and with the \textit{reasoning_tokens} value set to the default \textit{medium}, the model often returns a single word answer, ``Yes'' or ``No'', yet maintains high accuracy, see \autoref{t:cost}. 
OpenAI~\cite{o4minimodel} reports that the o-series models, including o4-mini, are trained to ``think longer'' before responding. 
The observation of single-token outputs can be explained by the hidden reasoning process executed in the backend during API calls. 

\PP{Performance differences across LLMs}
As discussed, without sufficient context all models struggle to identify \hpb precisely. 
When asked simply to decide whether a snippet is buggy or not, Llama‑4, two GPT models, and Claude 3.7 tend to answer ``Yes'', which yields high recall but low precision. Under this setup, only o4-mini achieves higher pairwise accuracy than the random baseline.
However, even provided with more information, the Rule configuration and the Patch configuration with the smaller models Llama-3.3-8b and gpt-4.1-nano still underperform. 
This suggests that current small models lack the depth of code understanding needed for reliable bug finding.
For the larger models, supplying a security coding rule produces strong and broadly consistent results.
Because patches are complex and contain extraneous detail, replacing the rule with a raw patch lowers pairwise accuracy for every model. The decline is smaller for the more robust frontier model like o4-mini.

\subsection{Overhead and Cost}
\label{ss:cost}

    \begin{table}[ht]
	\centering
        \footnotesize
        \resizebox{\columnwidth}{!}{%
        \begin{tabular}{ll|c|c|c|c|c|c}
\toprule
 
    & & \textbf{} & \textbf{} & \textbf{} &   \textbf{} &  \textbf{Rule}  & \textbf{Rule}\\ 
    & & \textbf{Basic} & \textbf{Patch} & \textbf{HuRule}  & \textbf{Rule}  & \textbf{w/o}  &     \textbf{+} \\
    & & \textbf{} & \textbf{}  & \textbf{}      & \textbf{} &  \textbf{CoT} &    \textbf{Patch}  \\ \hline 
    \multicolumn{2}{l|}{\textbf{Input Tokens}}      & 0.84 & 2.25     & 0.87    & 0.87  & 0.86 &    2.29\\ \hline\hline    
    \multirow{6}{*}{\textbf{Output Tokens}} & \graycc gpt-4.1-nano         & \graycc 1.27 & \graycc 0.93  & \graycc 0.75  & \graycc 0.80    & \graycc 0.004 & \graycc   0.86\\
  
     & Llama-3.3-8b           & 0.94 & 0.44 &  0.48   & 0.52    & 0.05 &    0.46 \\

     &\graycc Llama-3.3-70b         &\graycc 0.73 &\graycc  0.48 &\graycc  0.49    &\graycc 0.51     &\graycc 0.11 &\graycc    0.48\\

     &Llama-4-17b-128E      & 0.87 & 0.63 & 0.65       & 0.65 & 0.55 &   0.65 \\

     &\graycc Claude Sonnet 3.7    &\graycc 0.79 &\graycc 0.64 &\graycc 0.52 &\graycc  0.54     &\graycc 0.20  &\graycc     0.59 \\
     
     &o4-mini         & 0.50 & 0.29  & 0.27 &  0.28      & 0.001 &   0.27\\
\bottomrule
\end{tabular}
        }
        \caption{Average input and output tokens per prompt configuration (in thousands)}
        \label{t:cost}
    \end{table}

\begin{figure}[ht]
    \centering
    \includegraphics[width=\columnwidth]{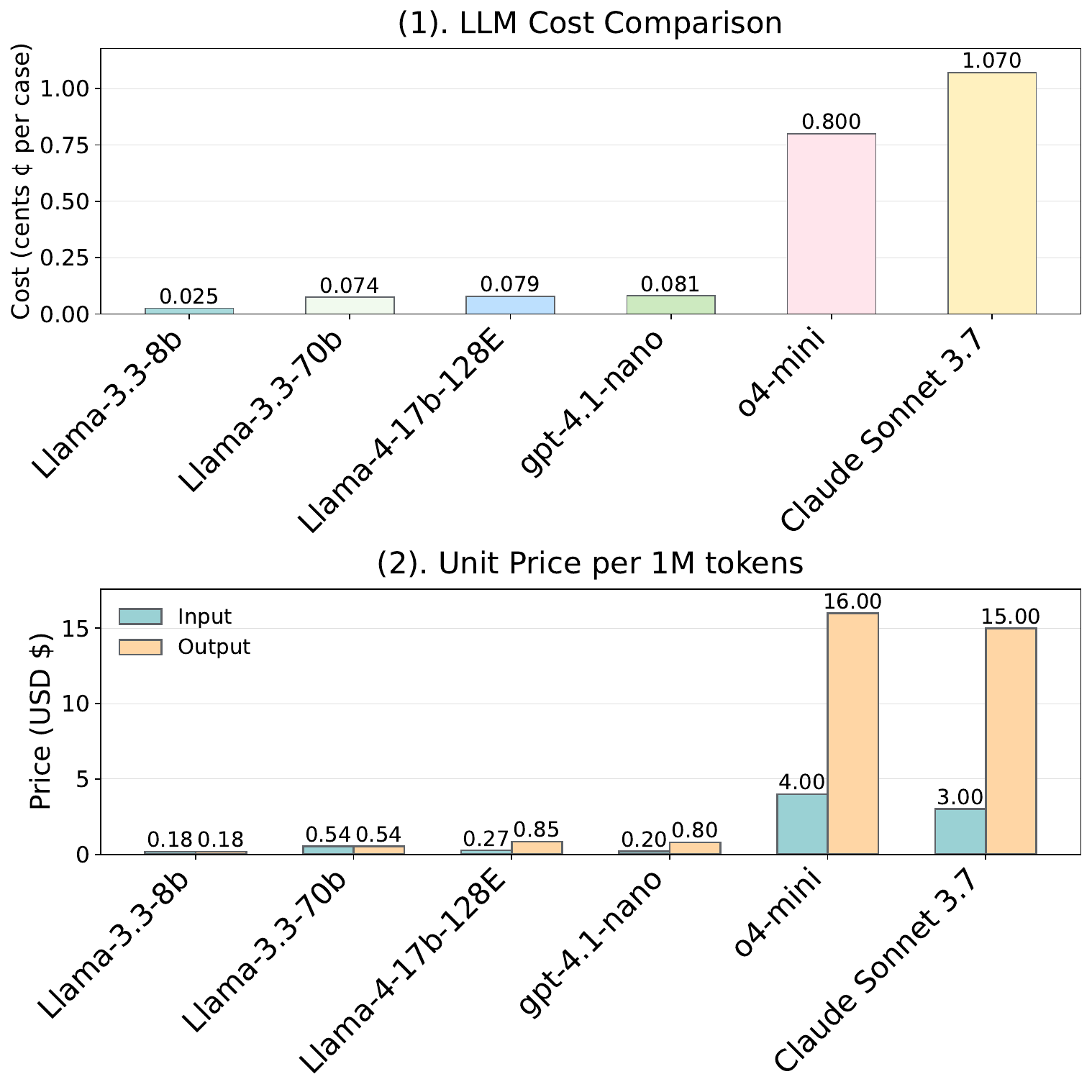}
    \caption{Average cost per detection (top) and unit price of LLMs (bottom) when using the Rule configuration prompt.}
    \label{fig:costpic}
\end{figure}

In this section, we evaluate the inference cost when varying the model and prompt configurations used by \sys. 
Because LLMs usually run as cloud services and providers use different hardware, raw latency is not a fair basis for comparison across models. 
Instead, we estimate the average input and output tokens for each model and configuration. Then, we estimate the cost using the prices by vendors.

\begin{table*}[!t]
	\centering
        \footnotesize
        \resizebox{\textwidth}{!}{%
        \begin{tabular}{c|c|l}
\toprule
 
  \textbf{TRId} & \textbf{\# Rules }    &  \textbf{Rule Template}  \\ \hline 
     
   1 & 54 &  The function \{TARGET\} may fail and return \{RETURN\_VALUE\}. Thus its return value should be checked before use to prevent \{IMPACT\} .  \\ \hline 
   \rowcolor{gray!20}
   2 & 57 &  Once \{TARGET\} succeeds, ensure that \{HANDLER\} is invoked in the subsequent error-handling path to prevent \{IMPACT\}. \\ \hline 
   3 &  9 & Use \{HANDLER\} instead of \{TARGET1\}+\{TARGET2\} to \{GOAL\}. \\ \hline
   \rowcolor{gray!20}
   4 & 2 & Memory allocated with \{TARGET\} must be freed with \{HANDLER\}, not kfree().  \\ \hline 
   
   \multirow{2}{*}{5} &  \multirow{2}{*}{3} & After calling \{TARGET\}, the refcount is incremented regardless of success or failure, so \{HANDLER\} must be invoked in every error-handling \\
    &  & path to prevent \{IMPACT\}. \\ \hline
    \rowcolor{gray!20}
   6 & 2 & Use \{HANDLER\} instead of \{TARGET\} when \{CONDITION\}, to prevent buffer overflow. \\ \hline
    
    \multirow{2}{*}{7} &  \multirow{2}{*}{1} &  Use \{HANDLER\} instead of \{TARGET\} for Ethernet‑address comparisons. This guarantees correct results and skips unnecessary bytewise \\
      &  & memory checks.	\\ \hline
    \rowcolor{gray!20}
    8 & 1 &  Use \{HANDLER\} instead of \{TARGET\} for delays under 20ms \\ \hline
    9 &  1 &  Release the  \{HANDLER\} before calling \{TARGET\} and reacquire it immediately afterward to prevent \{IMPACT\}. \\ \hline
    \rowcolor{gray!20}
    10 &  1 &  No need to call \{TARGET\} before destroying them with \{HANDLER\}, as it automatically drains them, thus avoiding unnecessary overhead. \\ \hline
    11 & 1 &  The \{TARGET\} function returns an `unsigned long` value instead of 'int'. Make sure the return value is put into a variable with unsigned long type. \\ \hline
    \rowcolor{gray!20}
    12  &  1 &  \{TARGET\} returns a negative value on failure, so the return check should be irq < 0 instead of irq == 0. \\ \hline
    \multirow{2}{*}{13} &  \multirow{2}{*}{1} & Instead of invoking \{TARGET1\} and \{TARGET2   \} separately, use the \{HANDLER\} helper for iomap operations. This ensures proper \\
       &  &  resource management and avoids potential issues. \\ \hline
     \rowcolor{gray!20}
     14 &  1 & The third parameter passed to core_link_read_dpcd() may remain uninitialized if the call fails, and since that variable might later be used by \\
     \rowcolor{gray!20}
   & & functions like core_link_write_dpcd(), it should be zero (e.g., with memset) before invoking core_link_read_dpcd() to prevent undefined behavior. \\ \hline
\bottomrule
\end{tabular}
        }
        \caption{135 unique rules classified by template rule Id (TRId); \# Rules: number of rules}
        \label{t:rules}
\end{table*}
    
%
\autoref{t:cost} shows the average number of tokens across all evaluated patches in the ground truth dataset. The number of input tokens reported in the first row is the same across all model configurations. When a seed patch is not provided, inputs are about 850 tokens, largely from the code piece under analysis. 
With a seed patch, the input size increases to around 2,250 tokens, an increase of roughly 160\% due to the commit message and expanded patched code context. 
This shows that supplying a patch as an example substantially increases the cost of \hpb detection.

The remaining rows show the average number of output tokens produced by each LLM in each setup. 
In the Basic setting, where accuracy is near random, outputs are the longest among all configurations. Without sufficient context and guidance, models produce verbose responses, thus increasing inference cost without any performance gain.
For a given model, average output length is often similar when the prompt allows for chain of thought (Patch, HuRule, Rule, Rule+Patch).
From shortest to longest average output length the order is o4-mini, Llama-3.3, Claude Sonnet 3.7, Llama‑4, and gpt-4.1-nano. In \autoref{fig:costpic}, we show the average cost per patch (top) when using the Rule configuration prompt. Even though o4-mini has the shortest average output length, it is drastically more expensive than all but Claude Sonnet 3.7 due to its high unit price~\cite{llamaprice, openaiprice, claudeprice}. Prices can be reduced by removing CoT, assuming performance is consistent. In practice, costs are also likely to be lower when performing large scale bug detection as the prompt template and rules are fixed, allowing for input caching~\cite{inputcach}. According to these results, Llama‑4 is currently the ideal choice for RPB detection, given its competitive performance with frontier models, but at a much cheaper cost.

\section{Large Scale RPB Detection }
\label{s:study}

In this section, we use \sys to identify novel \hpb in the Linux kernel. We describe the process to create a set of security coding rule templates and rules. Then, we re-analyze the performance of \sys on the ground truth dataset (\autoref{ss:gtcollection}) with respect to each rule template and discuss the cause of false positive and false negative findings. Finally, we share our findings and insights after applying \sys to the Linux kernel.

%
   

\PP{Creating rule templates} In \autoref{ss:gtcollection}, we manually reviewed 1,910 patches associated with state-of-the-art works and extracted 119 security coding rules that addressed API misuse or were misused code fragments. 39 of these rules were associated with a single patch that did not share a misuse pattern with other patches in the ground truth, but could still be used to identify novel \hpb. We further analyzed 1.06 million Linux kernel patches to extract other common recurring patterns that were not covered by the reviewed works. We hypothesize that bugs sharing the same recurring pattern will likely share similar patch descriptions and fixes, and thus can be grouped by these shared features.
%
%

%
Specifically, we first used a lightweight LLM, Llama-3.3-8b, to summarize each patch into a security coding rule. Then, we built a function–rule map by linking a function to any security coding rule that mentions it (many-to-many map). Using gpt4all~\cite{gpt4all}, we computed embeddings for each rule and applied a density-based clustering algorithm to group semantically similar rules in the embedding space. 
We selected the top 200 largest clusters representing approximately 17K patches. 
Through manual analysis and validation, we distilled these clusters into 34 representative rules whose violations could cause diverse impacts to kernel. 18 of these 34 rules overlap with the 119 rules derived from state-of-the-art works.
\autoref{t:rules} summarizes all these 135 unique rules distributed across 14 rule templates. Studying these templates highlights that \hpb have both security and non-security impacts. With \sys, once a specific security coding rule is available, similar \hpb can be identified across the program.
%
%


\subsection{Analyzing the Rule Templates}
\label{s:srules}

In this section, we will re-analyze performance of \sys using the ground-truth dataset of 80 rules, but split according to the rule templates. This analysis uses Llama-4 with the rule configuration prompt.

\autoref{t:accuracy_rules} shows that recent state-of-the-art works concentrate on bugs associated with rule templates 1, 2, and 5. These templates focus on memory leaks, reference count leaks, NULL pointer dereferences, and issues mainly related to security checks and error-handling. Rule template 3 covers issues related to performance, type misuse, concurrency, and reliability. Rule template 4 focuses on \cc{kfree()} issues.

\begin{table}[H]
	\centering
        \footnotesize
        \begin{tabular}{c|c|c|c|c|c}
\toprule
 
  \textbf{TRId} & \textbf{\# of rules} & \textbf{\# of cases}    &  $\mathcal{P}$ &  $\mathcal{R}$ & $\mathcal{PA}$  \\ \hline 
  \rowcolor{gray!20}
   1 & 34 & 194 & 97.1\% & 88.1\% & 85.4\% \\ \hline
   
   2 & 41 & 440 & 91.0\% & 83.0\% & 75.7\% \\ \hline
   \rowcolor{gray!20}
   3 & 1  & 17  & 100\%  & 100\%  & 100\% \\ \hline
   
   4 & 1  & 2   & 100\%  & 50\%   & 50\% \\ \hline
   \rowcolor{gray!20}
   5 & 3  & 117 & 87.7\% & 91.5\%  & 79.1\% \\ \hline\hline
   
  \textbf{Overall}  & \textbf{80} & \textbf{770} & \textbf{92.2\%} & \textbf{85.8\%} & \textbf{79.1\%} \\ \hline
   
\bottomrule
\end{tabular}
        \caption{Accuracy of \sys for different types per template rule Id (TRId).}
        \label{t:accuracy_rules}
    \end{table}

\sys achieves very high precision and pairwise accuracy on rule templates 1 and 3. These templates are easy to validate since it requires confirming if the safeguard check or required API call was made. Rule templates 2 and 5 exhibit richer semantics, including understanding error handling logic and control flow reasoning, which is marked by lower precision and pairwise accuracy.

To better understand where performance can be improved, we manually audited the false positive and false negative findings. From our analysis, we identify two main reasons for false positives:

\begin{enumerate}
    \item \textit{Limited in semantic and control flow reasoning capabilities:} Despite major advances in LLM reasoning, we still observe mistakes in code understanding. It sometimes misread execution order or failed to properly resolve code constraints. For example, when prompted to check that memory objects are released on error handling paths after a successful allocation, it flagged an error handling path that occurs before the allocation as a rule violation. Complex control flow involving \cc{goto} branches also results in similar mistakes by the LLM. 
    \item \textit{Semantic hallucinations:} In several instances, we found cases where the LLM flagged an incorrect, but semantically similar function. Specifically, when instructed to verify that the return value of \cc{clk_prepare} is checked in the code, it also commented on \cc{clk_prepare_enable}, a similarly named function. Such issues can be handled via deterministic post-processing of the flagged violations.
\end{enumerate}

Similarly, we identify four main reasons for false negative:
\begin{enumerate}
    \item \textit{Incomplete LLM analysis, especially for long code fragments:} The majority of false negatives are due to inherent limitations of the LLM.
    They cannot exhaustively enumerate all cases in the input code, particularly as the code length increases. 
    For example, when multiple error-handling paths or several target functions must be analyzed, the model often failed to cover all possibilities, even when prompted with chain-of-thought instructions requiring step-by-step reasoning.

    \item \textit{Difficulty with complex dataflow and constraint understanding:} When code involves complex control or dataflow, the LLM struggles to resolve the conditions correctly. 
    For instance, if an allocation function is conditionally invoked, its corresponding release function must also be conditionally invoked. The LLM often failed to capture such conditional relationships accurately.  

    \item \textit{Semantic hallucinations:} Hallucinations can also cause false negatives. 
    In a few cases, the model incorrectly assumed specific semantic behaviors for functions, and consequently decided that a security coding rule does not apply, when in fact it does.  

    \item \textit{Insufficient information under intra-procedural analysis:} In a few cases, accurate classification of the buggy (unpatched) version required additional context beyond the given function. 
    Without global or inter-procedural information, the model lacked sufficient semantics to make the correct decision, resulting in false negative reports.  
\end{enumerate}

\subsection{Debugging the Linux Kernel}
\label{s:ruleviolation}

Using the 135 rules and their associated APIs, \sys enumerated 148,664 candidate code pieces in the Linux kernel. These candidates were processed by Llama-4 with the rule configuration prompts resulting in 117M input tokens and producing 99M output tokens. In total, \sys flagged 22,568 potential rule violations. We randomly sampled 400 cases and, upon manual review, confirmed 246 as true violations, with varying potential security impacts if reachable.
Among the confirmed cases, 22 risk invalid pointer dereferences, 80 lead to memory, reference count, or other resource leaks, 85 result in type errors such as narrowing integer conversions, and the remainder involve miscellaneous issues such as performance degradation.
Although \sys has a non-negligible false positive rate in large-scale detection,
it remains manageable for manual review and is comparable to rates reported for state-of-the-art static analyzers~\cite{lin2023detecting,lu2019detecting,liu2021kubo,HERO}, which are often heavily tuned to specific patterns.
Our results reflect a single-shot run.
We expect reductions via multi-round prompting, cross-LLM validation, and tighter rule constraints with lightweight feasibility checks, which we leave to future work.
%
%

\PP{From reported cases to real issues}
The manual analysis of the 400 flagged violations revealed two reasons why a correct \sys detection of a rule violation might not be a true bug. 
First, despite a rule violation present in the scanned code piece, the program might remain correct due to deviations from the standard coding practice, such as module specific logic or data flow. 
For example, in \cc{init_urbs}, memory allocated by \cc{usb_alloc} is not released in the local error path, thus \sys flags this is a rule violation. Manual validation discovered that the callers of \cc{init_urbs} handle the failure and free the memory, so no leak occurs. 
Another example we found is when API behavior depends on an uncommon macro or input flag, so the rule does not always apply. In \cc{fs/ntfs3/super.c}, the return value of \cc{kmemdump} is unchecked, yet no issue arises because the call uses flag \cc{GFP_NOWARN}, which is designed to suppress handling for that error.

Second, some implementations might purposely violate the rule by design. 
Many module initialization functions, such as \cc{t_sdata_init}, omit error handling and return void or a constant value.
When asked, maintainers informed us that failures at such an early stage is unlikely, and if they do occur, the module simply exits.
Accurately recognizing these contexts and constraints requires stronger models or supplying richer contextual information to future analyses.

\PP{Responsible disclosure}
The sheer volume of findings makes manual individual review impractical and risks overwhelming maintainer even when we confirm the findings, but it remains critical to address the most severe issues. 
As mentioned previously, some true positive findings are currently non-issues, e.g., initialization functions that are robust to failures. 
At present, we prioritize findings by their associated rules and then manually assess their likely impacts to assign priority levels.
In our manual review of 400 cases, we found several \hpb that had already been discussed but not merged, for reasons such as maintainer prioritization, module specific semantics, or policies that only patch dynamically triggerable issues.
To reduce this manual effort, we are still working a tool that revalidates findings with more constraints to reducing false positives, and cross-references mailing list archives to determine whether an RPB has already been reported.

%

\PP{Limitations}
\sys offers practical advantages over many static analysis approaches. It can reason about semantic intent rather than relying on keyword matching and it does not rely on building a complex program analyzer for each unique bug type. For example, in module \cc{kirkwood_i2s}, the function \cc{kirkwood_i2s_dev_probe} calls \cc{clk_prepare_enable}. In one error path, it fails to invoke the corresponding unprepare function, which may lead to a resource leak problem. Detecting such issues via traditional program analysis requires field sensitive and flow sensitive analysis as well as substantial engineering effort. With \sys, this flow can be expressed as a single security coding rule. However, we recognize the limitation of intra-procedural reasoning. Issues like use-after-free that require cross function reasoning or thread interactions cannot currently be handled. Future work may solve this issue via an agentic workflow that alternates between a static analysis engine to extract relevant slices and an LLM to assess the extracted code. Additionally, findings are not guaranteed to be sound nor complete given unresolved issues in LLM performance. When asking a model, for example, to simply enumerate all the functions or variables in a file, it may return an incomplete list. This problem might resolve itself as models become more advanced.

\section{Related Work}
\label{s:relwk}

%

\PP{Bug Detection}
A large body of work~\cite{bai2019effective, lin2023detecting, tan2021detecting, liu2021kubo, lu2019detecting, HERO} has employed static analysis to detect semantic bugs in operating system kernels. 
These methods rely on manually crafted rules to capture recurring bug patterns, such as sleep-in-atomic-context~\cite{bai2019effective}, API post-handling errors~\cite{lin2023detecting}, reference-counting bugs~\cite{tan2021detecting}, missing-check bugs~\cite{lu2019detecting}, and disordered error handling~\cite{HERO}.
While these tools are effective for their targeted patterns, they require extensive engineering effort to build and maintain analyzers for each new bug class, and often lack generalizability beyond predefined templates.
Dynamic analysis techniques~\cite{kim2020hfl,ma2022printfuzz,jeong2019razzer} complement static methods by observing program behavior at runtime to detect the effects of bugs such as buffer overflows~\cite{ruwase2004practical,li2010practical,lhee2002type}, use-after-free errors~\cite{xu2015collision,wu2018fuze}, and NULL-pointer dereferences~\cite{machiry2017dr}. However, these techniques primarily focus on memory-related vulnerabilities and are inherently limited by incomplete execution path coverage and non-trivial performance overhead.
Complementing these program approaches, machine learning-based approaches have been explored, including unsupervised methods~\cite{ahmadi2021finding, yamaguchi2014modeling}, deep learning~\cite{li2018vuldeepecker, chakraborty2021deep}, and graph-based models~\cite{zhou2019devign, cheng2021deepwukong}. These methods offer improved adaptability and reduced manual effort, but may struggle with interpretability and require carefully curated training data.
Unlike prior work that relies on pattern-based static analyzers, dynamic instrumentation, or ML models trained on fixed vulnerability signatures, our approach treats bug detection as an adaptive reasoning task, utilizing LLMs’ ability to interpret diverse code semantics and historical bug contexts, without requiring static rules, handcrafted analyzers, or model retraining.

\ignore{
\noindent\textbf{API Misuse Detection.} 
API misuse detection~\cite{yun2016apisan,ren2020api,nielebock2020cooperative,li2021arbitrar,he2023one} has been extensively studied.
Some approaches~\cite{yun2016apisan,ren2020api} rely on semantic pattern mining, inferring correct API usage from code or documentation to identify violations. 
Others~\cite{nielebock2020cooperative,li2021arbitrar} adopt learning-based techniques: Nielebock et al.~\cite{nielebock2020cooperative} learn correction rules from developer patches, while Li et al.~\cite{li2021arbitrar} use active learning with symbolic execution to detect misuses with minimal developer input. He et al.~\cite{he2023one} present a large-scale empirical study showing that misunderstanding a single API can lead to hundreds of bugs.
}

\PP{LLM for Code Analysis}
LLMs have recently been explored for security-related code analysis tasks, such as vulnerability detection~\cite{ullah2024llms,yin2024multitask,gonccalves2024scope,kouliaridis2024assessing,lin2025large,liuexploring,chen2024witheredleaf}, malware analysis~\cite{patsakis2024assessing,huang2024exploring,zhou2025srdc}, and exploit generation~\cite{zhang2023well, dai2025comprehensive,nazzal2024promsec}. 
Several studies evaluate LLMs’ capabilities in vulnerability detection, revealing that off-the-shelf models often struggle with reasoning, localization, and consistency. 
For instance, Ullah et al.~\cite{ullah2024llms} report difficulties in handling bound checks and pointer operations. Yin et al.~\cite{yin2024multitask} observe that although LLMs such as CodeLlama perform well in descriptive tasks, their verbosity weakens their performance on other vulnerability related tasks. Other works highlight improvements through modest preprocessing~\cite{gonccalves2024scope}, perform evaluations across mobile and Android domains~\cite{kouliaridis2024assessing}, or perform systematic comparisons of model configurations and context lengths~\cite{lin2025large,liuexploring,chen2024witheredleaf}.  
Beyond evaluation, researchers have proposed techniques to enhance LLM-based vulnerability detection by enriching input context or adapting model architectures. 
Approaches in this space include path-sensitive representations~\cite{li2024llm}, graph-based encodings of control/data-flow~\cite{lu2024grace}, retrieval-augmented generation with structured vulnerability knowledge~\cite{du2024vul}, as well as domain-specific fine-tuning, adapter-based extensions, and lightweight models optimized for security tasks~\cite{shestov2024finetuning,luo2023wizardcoder,ferrag2023securefalcon,yang2024large,ma2024combining,ding2024vulnerability}. 
While prior work has explored LLMs for vulnerability detection, most studies evaluate models on constructed vulnerability benchmarks~\cite{lin2025large} or existing CVE datasets (e.g., NVD)~\cite{wang2025securemind}, rather than discovering new, real-world vulnerabilities.
The most closely related work is Li et al.~\cite{li2024enhancing}, which augments static analysis with LLMs to improve path sensitivity for detecting use-before-initialization (UBI) bugs. Our approach differs in three ways: \textit{(i) Method:} they enhance static analyzers, whereas we apply LLMs with security rules directly for detection; \textit{(ii) Scope:} they focus solely on UBI, while we cover multiple vulnerability classes; \textit{(iii) Findings:} they found 13 new bugs, whereas we identify thousands.

\PP{Code and Vulnerability Clone Detection}
Recent work on code and vulnerability clone detection has moved quickly away from token level similarity toward semantics aware matching. 
A common taxonomy classifies code clones into four types: Type 1 are exact copies except for whitespace and comments; Type 2 preserve structure but rename identifiers or literals; Type 3 are near misses with small edits, insertions, or deletions; and Type 4 are semantic equivalents that achieve similar behavior with different structures or APIs.
Siamese~\cite{ragkhitwetsagul2019siamese} performs incremental clone search by combining multiple code representations with query reduction and custom ranking, retrieving Type 1 to 3 clones. Gitor~\cite{shan2023gitor} retrieves clones via graph embeddings on a global code base graph linking code fragments to features. Tritor~\cite{zou2023tritor} detects semantic clones using a triads model that augments AST features with control and data flow.
Clone detection has also been applied to vulnerability discovery. VulDeePecker~\cite{li2018vuldeepecker} models “code gadgets,” i.e., semantically related lines, and learns their features with neural networks. MVP~\cite{xiao2020mvp} uses program slicing to extract vulnerability and patch signatures from ASTs and dependency graphs.  
However, these traditional clone based techniques are most effective for Type 1–3 similarity. For instance, MVP reports limitations in handling Type 4 clones, which require deeper semantic reasoning. Code Genome~\cite{kirat2024uncovering} extracts semantic fingerprint of the code and able to handle Type 4 clones, however, it is still limited to finding exact semantic code clones.
Our recurring patterns capture more than code cloning. 
They capture API misuse semantics at a high level using natural language representations rather than surface code tokens or structural similarity. Their callers may do unrelated tasks, so surrounding code often shares little syntax even as the same misuse pattern recurs. As a result, clone-based methods will struggle with these complex variants.

\ignore{
\subsection{LLM for code analysis}
\yue{To be updated}
LLMs have recently been explored for security-related code analysis tasks, such as vulnerability detection~\cite{}, malware analysis~\cite{patsakis2024assessing,huang2024exploring,zhou2025srdc}, and exploit generation~\cite{zhang2023well, dai2025comprehensive,nazzal2024promsec}. 

\noindent\textbf{Evaluating LLM's capability in vulnerability detection.}
While LLMs show promise for vulnerability detection tasks, recent studies reveal that off-the-shelf models often struggle with poor reasoning, hallucinations, and non-deterministic behavior.
Ullah et al.~\cite{ullah2024llms} conducted a comprehensive evaluation of LLMs’ ability to identify and reason about security-related bugs, showing that models often fail to localize and understand key constructs, particularly involving bounds checks, NULL checks, and arithmetic or pointer operations. 
Yin et al.~\cite{yin2024multitask} designed a multi-task benchmark using the Big-Vul dataset to assess detection, localization, assessment, and description tasks. While pre-trained models generally outperformed LLMs in detection, models such as CodeLlama and WizardCoder demonstrated stronger performance in assessment and description when given sufficient context. 
Gonçalves et al.~\cite{gonccalves2024scope} (SCoPE) showed that simple preprocessing—such as removing formatting noise—can lead to modest improvements in detection accuracy. 
Kouliaridis et al.~\cite{kouliaridis2024assessing} evaluated nine LLMs on Android OWASP Mobile Top 10 vulnerabilities, highlighting highly variable performance across domains. 
Similarly, Lin et al.~\cite{lin2025large} assessed numerous LLMs under different configurations, including model size, quantization, and context window, on C/C++ vulnerabilities from the Big-Vul dataset~\cite{fan2020ac}, finding significant variation in performance across models and languages. 
Liu et al.~\cite{liuexploring} compared ChatGPT against state-of-the-art vulnerability management tools, revealing strengths in summarization but weaknesses in complex reasoning tasks. 
Chen et al.~\cite{chen2024witheredleaf} conducted a systematic evaluation on detecting execution-injection bugs (EIBs), reporting that although GPT-4 shows potential, its limited precision and recall hinder its practical applicability.





\noindent\textbf{Enhancing LLMs for Vulnerability Detection.}  
Recent work has sought to improve LLM-based vulnerability detection by either enriching the input with more structured code context~\cite{li2024llm,lu2024grace,du2024vul} or enhancing model capabilities through fine-tuning and architectural adaptations~\cite{shestov2024finetuning,luo2023wizardcoder,ferrag2023securefalcon,yang2024large,ma2024combining,ding2024vulnerability}.
Specifically, several approaches focus on providing richer contextual information to LLMs. Li et al.~\cite{li2024llm} improved Java vulnerability detection by incorporating path-sensitive code context. Lu et al.~\cite{lu2024grace} proposed GRACE, which encodes control-flow and data-flow information via graph-structured representations to support more accurate LLM reasoning. Du et al.~\cite{du2024vul} introduced Vul-RAG, a retrieval-augmented generation framework that integrates multi-dimensional vulnerability knowledge—such as CWE definitions and prior patches—into prompts, significantly improving detection accuracy.
Complementary efforts enhance the models themselves. Shestov et al.~\cite{shestov2024finetuning} and Luo et al.~\cite{luo2023wizardcoder} applied domain-specific fine-tuning and instruction-tuning (Evol-Instruct) to improve code reasoning capabilities. Ferrag et al.~\cite{ferrag2023securefalcon} developed SecureFalcon, a compact 121M-parameter LLM optimized for efficient vulnerability classification. Yang et al.~\cite{yang2024large} employed adapter-based fine-tuning to support precise, line-level fault localization. Ma et al.~\cite{ma2024combining} combined fine-tuned LLMs with agent-based reasoning for smart contract vulnerability detection and explanation. Additionally, Ding et al.~\cite{ding2024vulnerability} curated a dedicated dataset for training and evaluating LLMs on vulnerability-related tasks.
}









\ignore{
\subsection{Others}

\noindent\textbf{Security Patch Identification \& Bug Prioritization.} 
A prominent line of work~\cite{zhou2017automated,wang2020machine, wang2021patchrnn, wu2020precisely,wu2022aware} focuses on identifying and prioritizing security patches. Zhou and Sharma~\cite{zhou2017automated} leverage natural language processing (NLP) on commit messages and bug reports to detect security-relevant changes in real time. Wang et al.~\cite{wang2020machine} apply machine learning to code changes and commit metadata to classify patches by vulnerability type. Building on this direction, PatchRNN~\cite{wang2021patchrnn} employs deep neural networks to discover ``secret'' security patches with low false positives. Complementarily, Wu et al.~\cite{wu2020precisely} propose SID, which uses symbolic rule comparison to assess the security impact of patches, enabling precise prioritization of critical fixes. Wu et al.~\cite{wu2022aware} further propose DIFFCVSS, combining static analysis and NLP to analyze the differential severity of vulnerabilities across Linux versions for OS-aware vulnerability triage.
}





\ignore{
\noindent\textbf{Duplicate Bug Report Detection and Analysis.} 
Another thread of research~\cite{mu2022depth,he2020duplicate,kukkar2020duplicate} addresses redundancy in bug reports. Mu et al.~\cite{mu2022depth} conduct a large-scale empirical study of duplicated Linux kernel bug reports, revealing their prevalence and root causes. To automate detection, He et al.~\cite{he2020duplicate} utilize a dual-channel convolutional neural network (CNN) to learn similarity between report pairs, while Kukkar et al.~\cite{kukkar2020duplicate} apply similar CNN-based methods to classify and filter duplicate reports, reducing triage effort.
}



\section{Conclusion}
\label{s:conclusion}
This paper introduces \sys, a rule-centric system that combines lightweight static analysis with LLM-guided reasoning to detect recurring pattern bugs (RPBs) from a single exemplar fix. 
Starting from a seed patch, \sys cleans and expands its context, extracts a standardized security coding rule that captures the root cause, required handling, expected impact, and scope constraints.
\sys further enumerates candidate sites across the code base, and then uses the LLM to judge rule compliance for each case.

Using a ground-truth dataset of more than 850 security patches drawn from 23 research works, \sys achieves 92.2\% precision and 79.1\% pairwise accuracy under the formatted-rule configuration. 
Applied at scale with 135 unique rules, \sys flagged over 22K potential violations in the Linux kernel. From 400 sampled reports, 246 were confirmed as previously unreported issues including invalid pointer dereferences, resource leaks, type errors, and performance defects. These results show that a rule-centric LLM approach can turn a single fix into project-wide improvements with high precision.


\bibliographystyle{sty/ACM-Reference-Format}
\bibliography{p,conf}

\clearpage
\appendix
\section{\sys Prompts}
\label{appendix}

	\begin{table}[h]
	\centering
        \footnotesize
        \begin{tabular}{Z{\columnwidth}}
\toprule
 
 \noindent \textbf{Prompt for generating the security coding rule}    \\ \hline 

   \noindent You are a software security expert. Your task is to generate one or more security coding rules from the given patch. Definition: A security coding rule is a concise statement that specifies
   the correct usage of a target API or code element. Violating this rule can
   introduce bugs or other issues.Use the templates below whenever they fit the patch. If none fit, write your own concise rule in a single sentence.\\
\noindent {[Security Coding Rule Templates]}\\
 \noindent  1. The function \{TARGET\} may fail and return \{ERR_RETURN_VALUE\}. Therefore,its return value must be checked before use to prevent \{IMPACT\}.\\
\noindent   2. Once \{TARGET\} succeeds, ensure that \{HANDLER\} is invoked in any 
   subsequent error handling path to prevent \{IMPACT\}.\\
\noindent   3. Use \{HANDLER\} instead of \{TARGET1\} + \{TARGET2\} to \{GOAL\}.\\
\noindent   {[END Security Coding Rule Templates]}\\
\noindent Output requirements:\\
\noindent1. Derive rules only from evidence in the patch, do not speculate.\\
\noindent2. Use identifiers as they appear in the patch.\\
\noindent3. If the patch addresses multiple independent issues, output multiple rules, one per line.\\
\noindent4. Do not include explanations or restate the patch.\\
\\
\noindent{[PATCH]}\\
\noindent\cc{\{PUT_PATCH_HERE\}}\\
\noindent{[END PATCH]}\\
\noindent Please provide the security coding rule or rules using the templates when possible. If no template fits, provide a concise custom rule.\\
 \\ \hline

\bottomrule
\end{tabular}
        \caption{Prompt for generating security coding rule. The placeholder \cc{\{PUT_PATCH_HERE\}} should be replaced with a patch.}
        \label{t:prompt_rule}
    \end{table}

	\begin{table*}[!ht]
	\centering
        \footnotesize
        \begin{tabular}{l|l}
\toprule
 
  \textbf{Prompt Setup} & \textbf{Prompt} \\ \hline 
    & 
You will be provided with a code. Your task is to analyze the code by following each code path and identify if the code contains\\
&any bugs, such as NULL dereference, memory leak, refcount leak, etc. \\
& [Code Snippets] \\
\textbf{Basic}& \cc{\{TARGET_CODE\}} \\
& [end] \\
&Analyze the code line by line and show the analyzing steps. Finally, respond with `YES' if there must be a violation of the coding\\
&rule, or `NO' otherwise. Do not assume other situations that are not appeared in the code or not mentioned in the coding rule. \\\hline

 & You will be provided with a patch. Afterwards, you will be shared with a code. Your task is to analyze the code by following\\
&each code path and identify if the code contains the same issue addressed by the patch. \\
&[Patch]\\
&\cc{\{PATCH\}}\\
\textbf{Patch}&[end]\\
&[Code Snippets]\\
&\cc{\{TARGET_CODE\}}\\
&[end]\\
&Analyze the code line by line and show the analyzing steps. Finally, respond with `YES' if there must be a violation of the coding\\
&rule, or `NO' otherwise. Do not assume other situations that are not appeared in the code or not mentioned in the coding rule. \\\hline

& You will be provided with a security coding rule. Afterwards, you will be shared with a code. Your task is to analyze the code\\
&by following each code path and identify if the code violate the security rule.\\
&[Security coding rule]\\
&\cc{\{RULE\}}\\
\textbf{Rule or HuRule}&[end]\\
&[Code Snippets]\\
&\cc{\{TARGET\_CODE\}}\\
&[end]\\
&Analyze the code line by line and show the analyzing steps. Finally, respond with `YES' if there must be a violation of the coding\\
&rule, or `NO' otherwise. Do not assume other situations that are not appeared in the code or not mentioned in the coding rule. \\\hline

&You will be provided with a security coding rule. Afterwards, you will be shared with a code. Your task is to analyze the code\\
&by following each code path and identify if the code violate the security rule.\\
&[Security coding rule]\\
&\cc{\{RULE\}}\\
\textbf{Rule w/o CoT}&[end]\\
&[Code Snippets]\\
&\cc{\{TARGET\_CODE\}}\\
&[end]\\
&Respond with `YES' if there must be a violation of the coding rule, or `NO' otherwise. Do not assume other situations that are not \\
&appeared in the code or not mentioned in the coding rule.\\ \hline

& You will be provided with a security coding rule, its related patch. Afterwards, you will be shared with a code. Your task is to\\
&analyze the code by following each code path and identify if the code violate the security rule.\\
&[Security coding rule]\\
&\cc{\{RULE\}}\\
 &[end]\\
&[Patch]\\
\textbf{Rule+Patch}&\cc{\{PATCH\}}\\
&[end]\\
&[Code Snippets]\\
&\cc{\{TARGET_CODE\}}\\
&[end]\\
&Analyze the code line by line and show the analyzing steps. Finally, respond with `YES' if there must be a violation of the coding\\
&rule, or `NO' otherwise. Do not assume other situations that are not appeared in the code or not mentioned in the coding rule. \\\hline

\bottomrule
\end{tabular}
        \caption{Prompt Configurations. Replace the placeholders \cc{\{ \}} with a security coding rule, a seed patch, or a target code fragment for analysis.}
        \label{t:prompts}
    \end{table*}
\end{document}